\documentclass[letterpaper]{article} 
\usepackage{aaai25}  
\usepackage{times}  
\usepackage{helvet}  
\usepackage{courier}  
\usepackage[hyphens]{url}  
\usepackage{graphicx} 
\usepackage{natbib}  
\usepackage{caption} 
\usepackage{algorithm}
\usepackage{algorithmic}

\usepackage{newfloat}
\usepackage{listings}
\DeclareCaptionStyle{ruled}{labelfont=normalfont,labelsep=colon,strut=off} 
\floatstyle{ruled}
\newfloat{listing}{tb}{lst}{}
\floatname{listing}{Listing}
\usepackage{longtable}
\usepackage{booktabs}

\usepackage[most]{tcolorbox}
\definecolor{mybackcolor}{HTML}{D9D9D9}
\tcbset{
  mypromptbox/.style={
    colframe=black,
    colback=mybackcolor,
    enhanced,
    rounded corners,
    boxrule=1pt
  }
}

\nocopyright

\title{Large Language Models Reflect Human Citation Patterns \\ with a Heightened Citation Bias}
\author{
Andres Algaba\textsuperscript{\rm 1}\thanks{Corresponding author. Data and code are available at \\ \color{blue} https://github.com/AndresAlgaba/LLM\_citation\_patterns.} \quad Carmen Mazijn\textsuperscript{\rm 1} \quad Vincent Holst\textsuperscript{\rm 1} \\ Floriano Tori\textsuperscript{\rm 1} \quad Sylvia Wenmackers\textsuperscript{\rm 2} \quad
Vincent Ginis\textsuperscript{\rm 1,3}
}
\affiliations{
    \textsuperscript{\rm 1}Vrije Universiteit Brussel, 
    \textsuperscript{\rm 2}KU Leuven, 
    \textsuperscript{\rm 3}Harvard University\\
\texttt{\{andres.algaba,carmen.mazijn,vincent.thorge.holst,floriano.tori\}@vub.be}\\
\texttt{sylvia.wenmackers@kuleuven.be} \quad 
\texttt{ginis@seas.harvard.edu}
}

\begin{document}

\maketitle

\begin{abstract}
Citation practices are crucial in shaping the structure of scientific knowledge, yet they are often influenced by contemporary norms and biases. The emergence of Large Language Models (LLMs) introduces a new dynamic to these practices. Interestingly, the characteristics and potential biases of references recommended by LLMs that entirely rely on their parametric knowledge, and not on search or retrieval-augmented generation, remain unexplored. Here, we analyze these characteristics in an experiment using a dataset from AAAI, NeurIPS, ICML, and ICLR, published after GPT-4's knowledge cut-off date. In our experiment, LLMs are tasked with suggesting scholarly references for the anonymized in-text citations within these papers. Our findings reveal a remarkable similarity between human and LLM citation patterns, but with a more pronounced high citation bias, which persists even after controlling for publication year, title length, number of authors, and venue. The results hold for both GPT-4, and the more capable models GPT-4o and Claude 3.5 where the papers are part of the training data. Additionally, we observe a large consistency between the characteristics of LLM's existing and non-existent generated references, indicating the model's internalization of citation patterns. By analyzing citation graphs, we show that the references recommended are embedded in the relevant citation context, suggesting an even deeper conceptual internalization of the citation networks. While LLMs can aid in citation generation, they may also amplify existing biases, such as the Matthew effect, and introduce new ones, potentially skewing scientific knowledge dissemination.
\end{abstract}

%

\section{Introduction}

\begin{figure*}[t!]%
\centering
\includegraphics[width=0.95\textwidth]{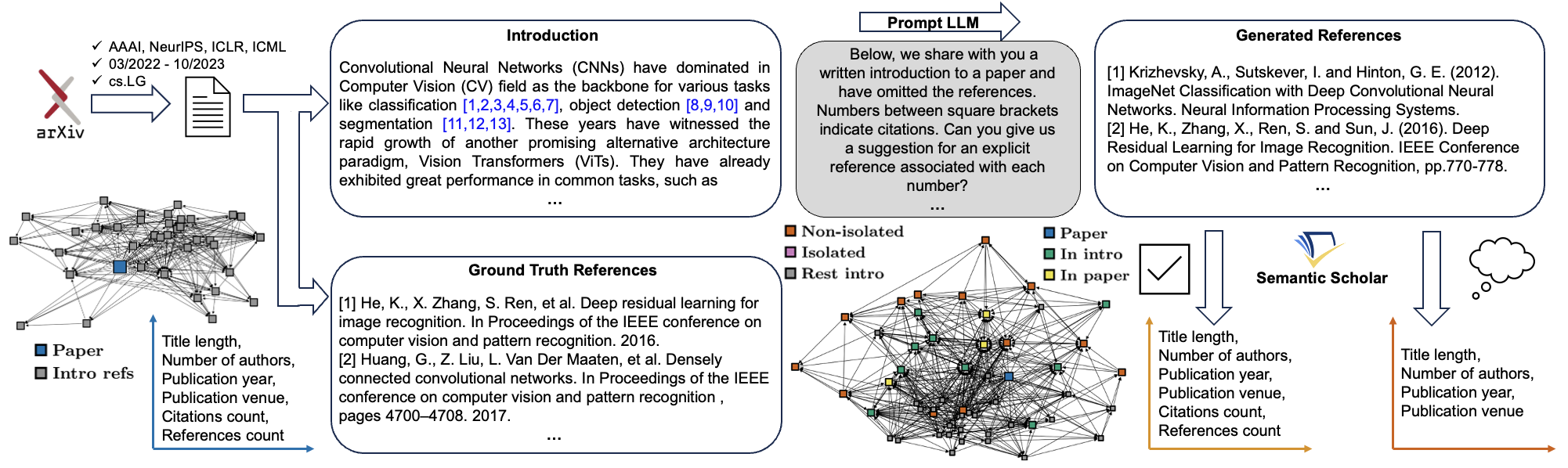}
\caption{\textbf{Overview of our experiment evaluating the characteristics and biases of LLM generated references, when tasked to suggest references for anonymized in-text citations.} We collect $166$ papers from the cs.LG category on arXiv which are published in the main tracks of AAAI, NeurIPS, ICML, and ICLR, and only appeared available online after GPT-4's knowledge cut-off date. We split the main content, which includes the author information, conference information, abstract, and introduction, from the ground truth references. GPT-4, GPT-4o and Claude 3.5 are prompted to generate suggestions of scholarly references for the anonymized in-text citations in the main content. We verify the existence of the generated references via Semantic Scholar and compare the characteristics, such as title length, publication year, venue, and number of authors, of the existing and non-existent generated references with the ground truth. For the existing generated references, we also compare additional characteristics, such as the number of citations and references, and analyze the properties of their citation networks.}
\label{fig:main_1}
\end{figure*}

Large Language Models (LLMs) have revolutionized natural language understanding and generation, driving scientific research forward by assisting in all steps of the scientific process, ranging from identifying research gaps to accelerating complex data analysis~\citep{boiko2023emergent,merchant2023scaling,romera2024mathematical,zheng2023large}. One particularly interesting application is the generation of suggestions for appropriate scholarly references~\citep{qureshi2023chatgpt,walters2023fabrication}. Yet, without the aid of web browsing or retrieval-augmented generation, these models rely entirely on their parametric knowledge encapsulated during their (pre-)training~\citep{brown2020language,bubeck2023sparks,kaddour2023challenges,wei2022emergent}. Our research focuses on this intrinsic citation behavior of GPT-4, exploring how the model recommends references based on its training data, and highlighting the potential biases that arise from this internalized knowledge~\citep{acerbi2023large,manerba2023social}.

Biases in citation practices have long been a subject of scrutiny in the scientific community~\citep{fortunato2018science,smith2012impact}. Common biases include a preference for recent publications~\citep{bornmann2008citation}, shorter titles~\citep{letchford2015advantage}, and high-profile publication venues~\citep{lawrence2003politics}. Moreover, a well-documented phenomenon is the ``Matthew effect,'' where highly cited papers tend to accumulate even more citations~\citep{wang2014unpacking}. The number of authors on a paper can also influence its likelihood of being cited, with solo and small author groups often cited less frequently than larger teams~\citep{gazni2011investigating}. By examining how these biases manifest in LLM generated references, we aim to uncover the underlying patterns and their potential to amplify existing biases and introduce new ones, potentially skewing scientific knowledge dissemination.

In our experiment, we let GPT-4, GPT-4o and Claude 3.5 suggest scholarly references for anonymized in-text citations within a paper and compare the characteristics and citation networks of the LLM generated references against the ground truth. We provide a comprehensive analysis of $166$ papers which are published in the main tracks of AAAI, NeurIPS, ICML, and ICLR, encompassing 3,066 references in total. All the papers are only first available online on arXiv after GPT-4-0613's knowledge cut-off date and belong to the cs.LG category. While this experimental setup may not fully reflect real-world usage of LLMs for citation generation, which often involves more interactivity and reliance on external data sources, it provides a controlled laboratory setting to assess the parametric knowledge and inherent biases of LLMs. Furthermore, our focused sample of papers ensures a homogeneous dataset, which allows us to minimize confounding factors that could arise from cross-disciplinary differences in citation practices.

Our setting differs from previous work which either let the LLM generate short papers or literature reviews, or is prompted for the most important papers on a certain topic~\citep{walters2023fabrication}. We argue that these methods are more susceptible to the LLM's memorization capabilities~\citep{Chen_Lin_Han_Sun_2024,kadavath2022language}. Moreover, the evaluation of the suggested references mostly focuses on their existence, bibliometric accuracy, or qualitative judgement by domain experts~\citep{qureshi2023chatgpt}. Finally, another strand of the literature focuses on improving LLMs via search and retrieval-augmented generation~\citep{lewis2020retrieval} or to reduce their hallucination rate via self-consistency~\citep{agrawal-etal-2024-language} to enhance their capabilities in systematic literature reviews~\citep{susnjak2024automating}.

In our experiment, we find that GPT-4 exhibits strong preferences for highly cited papers, which persists even after controlling for multiple confounding factors such as publication year, title length, venue, and number of authors. Additionally, we observe a large consistency between GPT-4's existing and non-existent generated references, indicating the model's internalization of citation patterns. The same results hold for the more capable models GPT-4o and Claude 3.5 where the papers are part of the training data. By analyzing citation graphs, we show that the references recommended by GPT-4 are embedded in the relevant citation context, suggesting an even deeper conceptual internalization of the citation networks. While LLMs can aid in citation generation, we find they may also amplify existing biases and introduce new ones, potentially skewing scientific discourse. Our results underscore the need for identifying the model's biases and for developing balanced methods to interact with LLMs in general~\citep{navigli2023biases}.

\section{Generating Citations with LLMs}
Our data consists of $166$ papers published at AAAI ($25$), NeurIPS ($72$), ICML ($38$), and ICLR ($31$) for a total of $3,066$ references. Our data collection process is depicted in Figure~\ref{fig:main_1} (see Appendix \ref{sec:data} for more details) and begins by retrieving all the relevant papers from arXiv, focusing on those within the machine learning category (cs.LG) and posted between March 2022 and October 2023 (after GPT-4-0613's knowledge cut-off date). The papers are verified on Semantic Scholar where we store additional metadata, such as all the reference titles with corresponding Semantic Scholar IDs to construct the citation networks (see Appendix Table \ref{tab:paper_details} for a full list of all the included papers).

We split the main content, which includes the author information, conference information, abstract, and introduction, from the ground truth references. Next, we prompt GPT-4, GPT-4o and Claude 3.5 as follows:
\begin{tcolorbox}[mypromptbox]
\begin{center}
    Below, we share with you a written introduction to a paper and have omitted the references. Numbers between square brackets indicate citations. Can you give us a suggestion for an explicit reference associated with each number? Do not return anything except the citation number between square brackets and the corresponding reference. \\
    === \\
     \textbf{\lbrack main content\rbrack}
\end{center}
\end{tcolorbox}
We then post-process the responses to extract the title, venue, publication year, author names, and number of authors for each generated reference (see Appendix \ref{sec:data} for more details). To assess the robustness of this approach, we repeat this ``vanilla'' approach three to five times for all $166$ papers.

A well-known issue in text generation by LLMs are hallucinations or confabulations, which refer to generated content that is nonsensical or untruthful in relation to certain sources, i.e., factual mistakes about historical events~\citep{zhang2023siren}. This is particularly problematic for the generation of scholarly references, as LLMs can fabricate references that do not exist or introduce subtle errors, making it impossible to retrieve the actual references~\citep{walters2023fabrication}. There are two main approaches to verify the existence of LLM-generated references: one involves asking additional questions to the LLM to verify its self-consistency~\citep{agrawal-etal-2024-language}, and the second approach utilizes external databases to verify a reference's existence~\citep{fabiano2024optimize}. In our experiment, we opt for the latter and determine via title and author names matching with Semantic Scholar entries whether the generated references exist (see Appendix \ref{sec:data} for more details). Finally, we also build on our ``vanilla'' approach, by introducing an ``iterative'' approach where we continue to prompt GPT-4 after having indicated which generated references do not exist and ask to replace those with existing ones (see Appendix \ref{sec:data} for more details). The previously existing generated and the newly generated references are then merged.

In Table~\ref{tab:extended_1}, we report the GPT-4 summary statistics for each of the five vanilla (and iterative: results between brackets) runs. On average, $65\%$ ($86\%$) of the generated references match with an entry in Semantic Scholar, while $13\%$ ($14\%$) and $17\%$ ($20\%$) of them appear in the introduction or paper itself, respectively. We further show that about $7\%$ ($7\%$) of the generated and ground truth references match pairwise. However, this number grows to $13\%$ ($14\%$) if we only consider the uniquely identifiable references (i.e., omitting references included in \textit{\lbrack 4--8\rbrack} as there is no one-to-one correspondence). In Appendix Table \ref{tab:overlap}, we show that the average overlap between generated sets is $17\%$.

\begin{table}[t!]
\centering
\setlength{\tabcolsep}{1mm}
\begin{tabular}{cccccc}
\textbf{Vanilla (Iterative)} & Run 1 & Run 2 & Run 3 & Run 4 & Run 5 \\ 
\hline
 \addlinespace
Existence & 64.3 & 63.3 & 62.8 & 64.2 & 67.6 \\ 
 & (87.0) & (85.5) & (88.0) & (86.8) & (86.3) \\
 \addlinespace
Cited in paper & 17.5 & 17.1 & 15.7 & 16.8 & 18.0 \\ 
 & (20.0) & (20.1) & (18.4) & (19.2) & (20.8) \\
  \addlinespace
Cited in introduction & 13.4 & 13.2 & 12.2 & 12.9 & 13.9 \\ 
 & (14.5) & (15.0) & (13.5) & (14.3) & (15.3) \\ 
  \addlinespace
Pairwise Match (PM) & 7.0 & 7.2 & 6.3 & 6.9 & 6.7 \\ 
for all references & (7.1) & (7.3) & (6.6) & (7.0) & (7.1) \\ 
 \addlinespace
PM for uniquely & 12.5 & 13.7 & 12.5 & 13.7 & 13.3 \\ 
identifiable references & (12.5) & (14.1) & (12.9) & (13.7) & (14.0) \\
\end{tabular}
\caption{\textbf{Summary statistics of GPT-4 generated references.} (in \% with respect to total number of references).}
\label{tab:extended_1}
\end{table}

\begin{figure*}[t!]%
\centering
\includegraphics[width=0.8\textwidth]{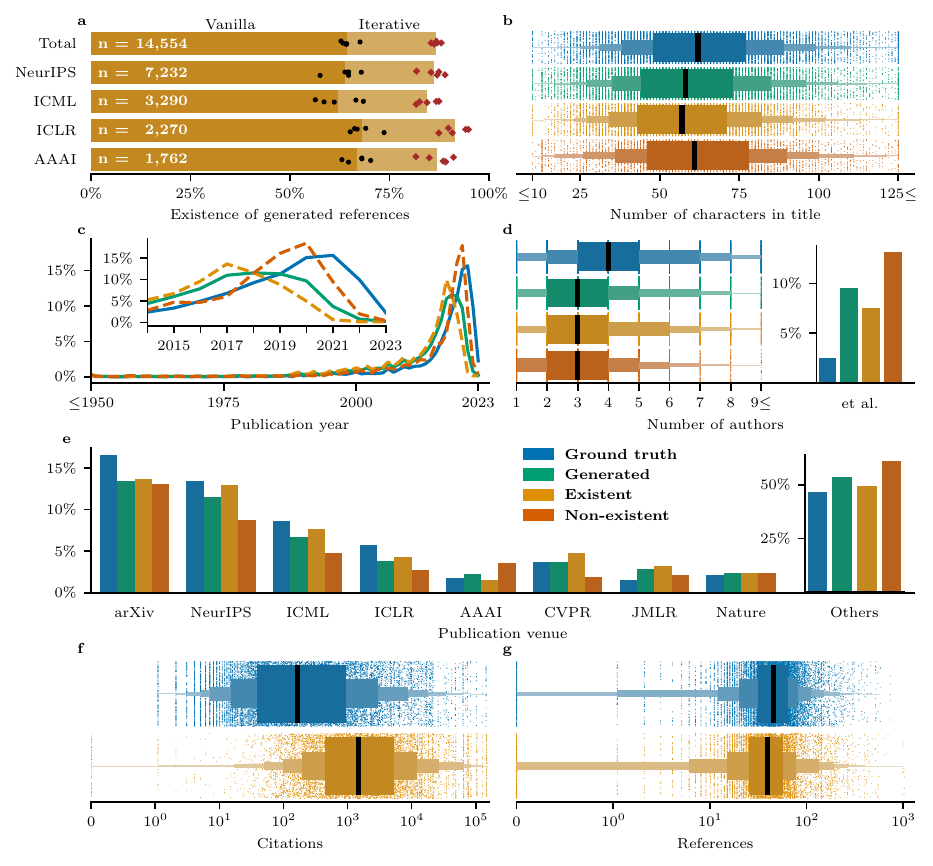}
\caption{\textbf{Properties of the ground truth and GPT-4 generated introduction references for the vanilla strategy.} This figure displays the properties of the ground truth ($n=14,554$, in blue) and GPT-4 generated references ($n=14,554$, in green), further subdividing the generated references into existing ($n=9,376$, in orange) and non-existent ($n=5,178$, in red), from the original data sources of five runs for the vanilla strategy with GPT-4. \textbf{a,} The average percentage of existing generated references in total ($64.4\%$) and for each publication venue under the vanilla and iterative strategy, with dots representing the percentage for each of the five runs with GPT-4. \textbf{b,} The distribution of the number of characters in the title shows some differences between the ground truth (median $62$) and generated (median $58$) references with the non-existent being slightly longer and with a larger variance compared to the existing generations. \textbf{c,} The distribution over time reveals that ground truth references are relatively more recent than generated references, with most references post-$2010$. The temporal distribution of the non-existent generated references aligns more with the ground truth than the existing generated references. \textbf{d,} The distribution of the number of authors demonstrates a disparity between the ground truth and generated references, having median values of three and two, respectively. However, GPT-4 more often generates ``et al.'' which does not allow for an exact computation, especially for the non-existent references. \textbf{e,} The distribution of publication venues shows that for most venues the ground truth has the highest relative representation, followed closely by existing references. The non-existent references deviate more from the ground truth as the proportion of ``Others'' is substantially larger. \textbf{f,} The distributions of citations for ground truth and existing generated references reveal a substantial citation bias in the generated references with a difference in median citations of $1,326$. \textbf{g,} Finally, the distribution of references shows that ground truth references cite slightly more papers than the existing generated references with a median difference in median references of $6$.}
\label{fig:main_2}
\end{figure*}

\section{Reflecting Human Citation Patterns}
Figure~\ref{fig:main_2} displays the characteristics of the ground truth and GPT-4 generated references, and separately the characteristics of the generated references which match with a Semantic Scholar entry, and those which do not exist according to this database. Overall, we observe a remarkable similarity between human and LLM citation patterns and a large consistency between GPT-4's existing and non-existent generated references, indicating the model's internalization of citation patterns. In Appendix Figure~\ref{fig:extended_1}, we also show that the newly generated papers from the ``iterative'' approach show nearly identical distributions.

The distributions of the title lengths show that existing generated reference titles tend to be the shortest, while non-existent generated reference titles are more similar in length to the ground truth, which indicates a learned pattern. Overall, the first effect dominates, so the average is skewed to shorter titles for generated references (Figure~\ref{fig:main_2}b). The temporal analysis reveals a similar pattern where non-existent generated references follow a distribution that is more similar to the ground truth than the existent ones (Figure~\ref{fig:main_2}c).

The distribution of the number of authors highlights a notable difference, with ground truth references typically involving three authors versus two for generated references, though the frequent use of ``et al.'' in the generated references complicates exact author counts (Figure~\ref{fig:main_2}d). To further examine the potential impact of the ``et al.'' problem, we only consider the existing generated references and their ground truth counterpart in Appendix Figure~\ref{fig:extended_2}. There, we compare the characteristics of the references between two data sources, namely the original source (the paper or GPT generation) and the available information on Semantic Scholar. The similarity between the distributions of all characteristics shows that the data source has no impact and ``et al.'' does not cause this observation.

The publication venue distributions show that for most venues the ground truth has the highest relative representation, followed closely by existing generated references, with non-existent generated references displaying the largest proportion of ``Others'' (Figure~\ref{fig:main_2}e). In Appendix Figure~\ref{fig:extended_4}, we observe that the distributions of publication venues for both ground truth and generated references are very similar across the various conferences, i.e., AAAI, NeurIPS, ICML, and ICLR. The pairwise transition matrix from ground truth to generated publication venues at the reference level indicates a large overall agreement, but with a strong preference in GPT-4 generated references for arXiv, NeurIPS, and ``Others'' in the case of disagreement. The preference for NeurIPS may be due to the relatively large number of NeurIPS papers in our sample and the large share of arXiv and ``Others'' points to favoring a wider array of venues which may potentially dilute the perceived relevance of key conferences. Finally, the scatter plot affirms the strong pairwise correlation between the ground truth and generated references to the top conferences at the individual paper level.

Most prominently, we observe a significant citation bias in the existing generated references, which have a median citation count of $1,326$ higher than ground truth references (Figure~\ref{fig:main_2}f). In Appendix Figure~\ref{fig:extended_3}, we compare the characteristics for the corresponding ground truth references of existing and non-existent references, and for the existing references which also appear in the paper itself. We observe that the ground truth papers which correspond to existing references that appear in the paper itself have by far the most citations, followed by the existing references, and the ground truth papers corresponding to non-existent references have the lowest numbers of citations. These findings further indicate the tendency for GPT-4 to more easily generate references to highly cited papers. Finally, the distribution of references indicates that ground truth references cite slightly more papers than existing generated references (Figure~\ref{fig:main_2}g).

In Appendix Figures \ref{fig:extended_gpt_4o} and \ref{fig:extended_claude_35} and Table \ref{tab:extended_2}, we find similar results for three GPT-4o and Claude 3.5 runs, but with a higher existence rate which may be due to the models' capabilities or the papers being part of the training data.

\begin{figure*}[t!]%
\centering
\includegraphics[width=0.7\textwidth]{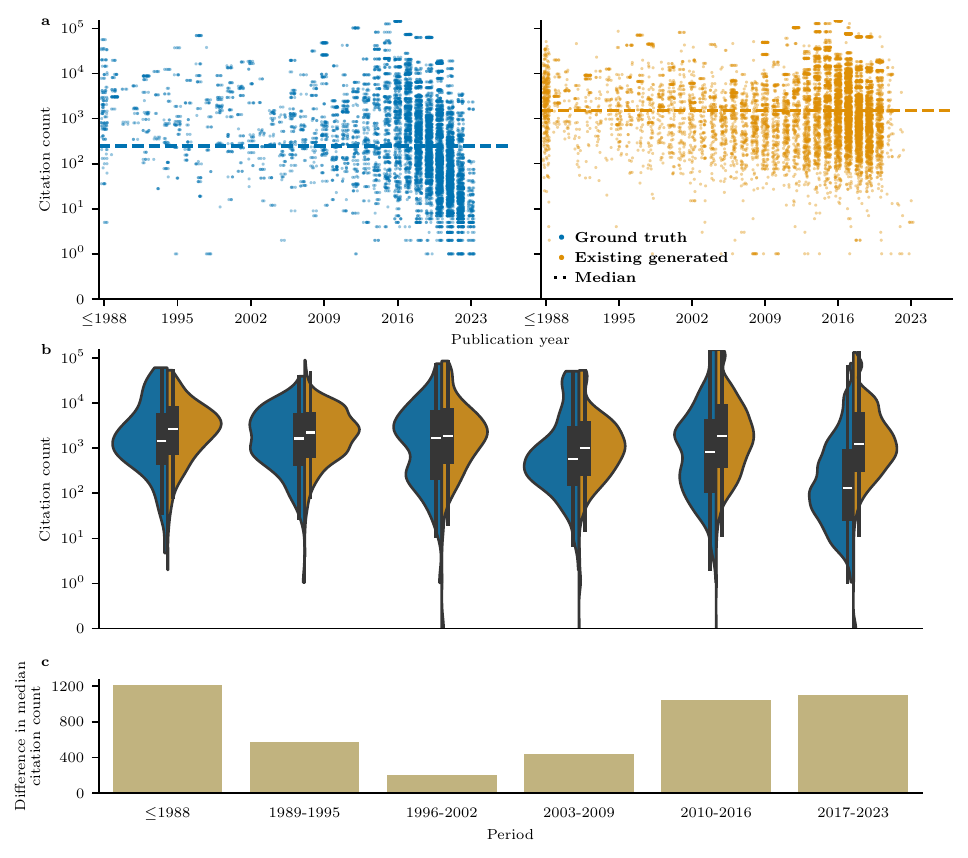}
\caption{\textbf{The citation bias in existing GPT-4 generated references is not due to the recency of ground truth references.} This figure shows that the existing GPT-4 generated references ($n=9,376$, in orange) consistently exhibit a higher citation count compared to their corresponding ground truth ($n=9,376$, in blue) across subperiods. \textbf{a,}~The citation counts across time for the ground truth and existing generated references reveal that the most recent references have a relatively low number of citations. The difference in median citations between the existing generated references and their corresponding ground truth references is 1,257. Since the ground truth references are relatively more recent compared to the existing generated references, we examine whether the observed citation bias is related to the recency of ground truth references. \textbf{b,} The distributions of citations by subperiod reveal that the existing generated references consistently exhibit a higher citation count than their corresponding ground truth counterparts. \textbf{c,} The difference in median citations is most pronounced in the early and late subperiods, i.e., $\leq 1988$, 2010-2016, and 2017-2023.}
\label{fig:main_3}
\end{figure*}

\section{Heightened Citation Bias}
Figure~\ref{fig:main_3} demonstrates that the citation bias observed in GPT-4 generated references is not merely a consequence of the recency of ground truth references. Specifically, the existing generated references show consistently higher citation counts compared to their ground truth counterparts across various subperiods. Figure~\ref{fig:main_3}a illustrates that ground truth references, particularly the most recent ones, tend to have lower citation counts. Despite the ground truth references being more recent on average, the citation counts of existing generated references remain significantly higher. Figure~\ref{fig:main_3}b further breaks down the citation distributions by subperiods, reaffirming that generated references consistently have higher citation counts than their corresponding ground truth references. Figure~\ref{fig:main_3}c highlights that this citation discrepancy is most pronounced in both the earliest ($\leq$1988) and the most recent (2010-2016 and 2017-2023) subperiods, indicating that the citation bias persists across different time frames.

In Appendix Figure~\ref{fig:main_4}, we find that the heightened citation bias in generated references remains also after controlling for other possible confounding factors, such as title length, number of authors, and publication venue. In Appendix Figure~\ref{fig:extended_5} and Appendix Figure~\ref{fig:extended_6}, we confirm that our findings are robust for the influential citation count which can be retrieved from Semantic Scholar~\cite{valenzuela2015influential}. This consistency across multiple factors underscores the inherent bias of LLMs towards generating references to highly cited papers, irrespective of other characteristics of the references.

\begin{figure*}
\centering
\includegraphics[width=0.85\textwidth]{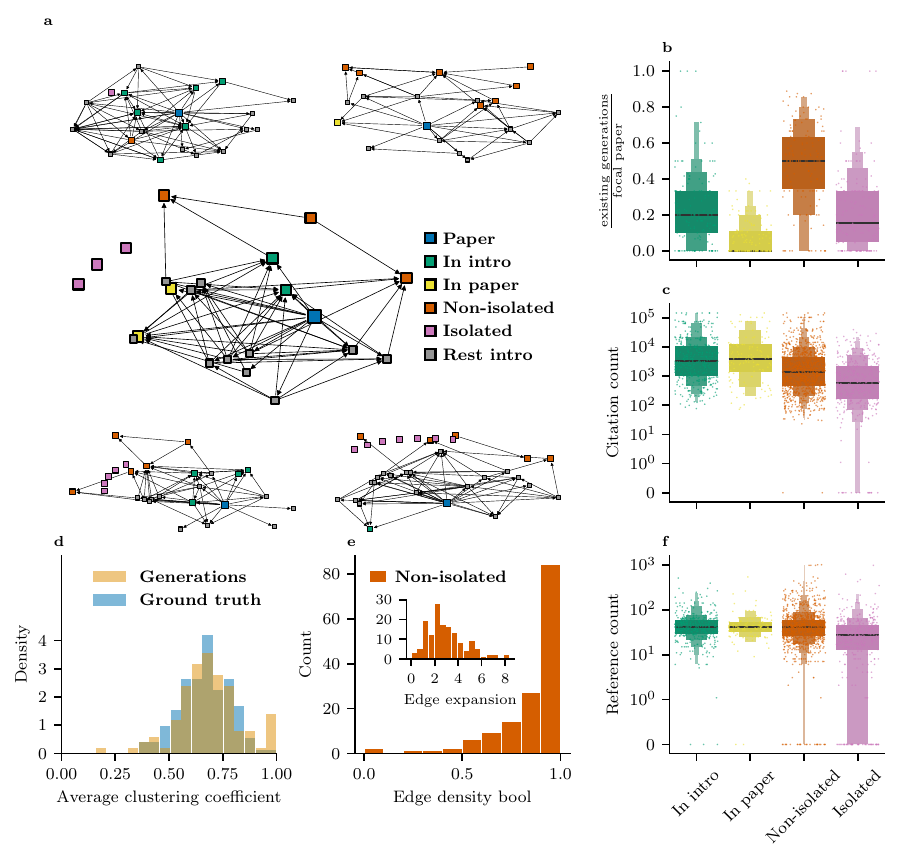}
\caption{\textbf{The GPT-4 generated references display similar citation network properties as the ground truth references but with a heightened citation bias.} This figure displays how the existing GPT-4 generated references ($n=2945$, first run of vanilla strategy) are embedded in the citation network of the focal papers ($m=166$ in total). \textbf{a,} We depict the connections between the focal paper, the ground truth references, and the existing generated references by showing the underlying citation graphs. An arrow from A to B indicates that A cites B. We identify the focal paper (in blue), generated references that appear in the introduction (in green) or in the paper (in yellow), generated references that are linked to ground truth or other generated references (in orange), generated references that are completely isolated (in purple), and ground truth references that are not cited by GPT-4 (in gray). \textbf{b,} The majority of generated references does not appear in the introduction or paper itself, but is somehow connected to the ground truth references as only a small fraction of generated references is completely isolated. \textbf{c,} The heightened citation bias is most emphasized for generated references that appear in the introduction or the paper, with isolated generated references having the lowest number of citations. \textbf{d,} The normalized average clustering coefficients of the ground truth (green and grey nodes) and the existing generated references (green, yellow, orange, and purple nodes) indicate that GPT-4's internalization of citation patterns extends to citation network properties. The clustering coefficient for a node A is given by $\frac{\# \text{triangles through A}}{\# \text{possible triangles through A}}$. The average is computed across the coefficients of all nodes in the respective graph (excl. nodes with coefficient zero) and indicates the tendency of the respective references to appear in clusters. \textbf{e,} The non-isolated generated references are tightly connected to the ground truth references, both on an individual level (Boolean edge density) as well as an aggregate level (edge expansion). The Boolean edge density is the fraction of non-isolated generations (orange nodes) that are connected to at least one ground truth reference (green and grey nodes) per focal paper. The edge expansion between those two sets is defined as the number of edges between the two sets divided by the smallest set size. \textbf{f,} The number of references is similar across all categories, except for the isolated generated references which have substantially less references.}
\label{fig:main_5}
\end{figure*}

\section{LLMs and Human Citation Networks}
Figure~\ref{fig:main_5} displays the properties of the ground truth and GPT-4 generated citation networks. In Figure~\ref{fig:main_5}a, we identify the focal paper (in blue), generated references that appear in the introduction (in green) or later in the paper (in yellow), generated references that are linked to ground truth or other generated references (in orange), generated references that are completely isolated (in purple), and ground truth references that are not cited by GPT-4 (in gray). The majority of generated references ($>50\%$) is non-isolated, i.e., linked to the ground truth or generated references but not present in the focal paper itself, followed by a substantial amount of generated references appearing in the introduction and only a small fraction that do not appear in the introduction but still within the focal paper (Figure~\ref{fig:main_5}b). The remainder of generated references is completely isolated from the citation network. If GPT-4 did not pick up on human citation patterns, the generated citation network would resemble a random network containing only isolated citations. The heightened citation bias is also most pronounced for references that appear within the introduction or paper, with isolated generated references having the lowest number of citations (Figure~\ref{fig:main_5}c). This finding further indicates the tendency for GPT-4 to more easily identify and generate references to highly cited papers. The number of references is similar across all categories, except for the isolated generated references which have substantially less references (Figure~\ref{fig:main_5}f).

The normalized average clustering coefficients of the ground truth (green and grey nodes) and the existing generated references (green, yellow, orange, and purple nodes) indicate that GPT-4’s internalization of citation patterns extends to citation network properties (Figure~\ref{fig:main_5}d). This internalization is also reflected by the tight connection between the non-isolated generated and ground truth references. The connection appears on an individual level as measured by the Boolean edge density, as well as on the aggregate level as measured by the edge expansion. For instance, in the central graph shown in Figure~\ref{fig:main_5}a, a Boolean edge density of $\frac{2}{3}$ suggests one non-isolated generated reference links only within its group, while an edge expansion of $2\frac{1}{3}$ indicates strong connections between the other two non-isolated generated references and the actual ground truth references. So, we can exclude the possibility of GPT-4 generating suggestions of scholarly references that are connected to each other but move further and further away from actual content of the introduction. Regardless, three of the four categories (green, yellow and orange) are well embedded in the given citation context. It reflects how tight the connection between the non-isolated generations to the ground truth references is and the deeper conceptual internalization of the citation networks.

\section{Discussion}
We present an experiment to explore the intrinsic citation behavior of LLMs and their potential biases when generating scholarly references. Whereas, previous work focuses on LLMs generating short papers or literature reviews~\citep{qureshi2023chatgpt,walters2023fabrication}, we let GPT-4, GPT-4o and Claude 3.5 generate suggestions of scholarly references for anonymized in-text citations. Importantly, we do not enhance the LLM through search and retrieval-augmented generation, but evaluate the model's internalization of citation patterns in its parametric knowledge obtained during training. While, our experimental setup may not fully reflect real-world usage of LLMs for citation generation, which often involves more interactivity and reliance on external data sources, it provides a controlled laboratory setting to assess the parametric knowledge and inherent biases of LLMs.

Our findings are significant as they represent a first step towards understanding the real-world impact of LLMs in scientific research~\citep{boiko2023emergent,lu2024aiscientist,zheng2023large}. By highlighting the heightened citation bias in LLM generated references, we demonstrate the models' tendency to favor highly cited papers, which could exacerbate existing biases in scientific discourse. This evaluation moves beyond more traditional LLM benchmarks~\citep{srivastava2022beyond}, emphasizing the practical implications of deploying these models in academic contexts~\citep{jimenez2024swebench}. The results suggest that while LLMs have the potential to streamline various aspects of research, careful consideration is needed to mitigate the amplification of biases, such as the ``Matthew effect.''

One plausible hypothesis for the heightened citation bias observed in LLMs is the increased frequency of citations to heavily cited papers within the model's training data. This prevalence makes these references more likely to be generated accurately and recognized as existent. Additionally, such biases may stem from generic training effects, where models preferentially learn patterns that are more common in the data, leading to biases towards shorter titles, more heavily cited, and slightly less recent works~\citep{kandpal2023large}. These tendencies may persist despite improvements in data quantity or model sophistication as indicated by our experiments with GPT-4o and Claude 3.5.

We develop and open-source an extensible, automated pipeline to systematically analyze the references generated by LLMs. Although our methodology is robust, it is not without limitations. The use of simple prompts and the zero-shot setting~\citep{kojima2022large} aims to minimize bias in the generation process, but this simplicity might not capture the full spectrum of potential LLM capabilities. There are numerous alternative approaches and prompt designs that future research can explore to enhance the accuracy and relevance of generated references~\citep{wang2022self,wei2022chain,yao2024tree}. However, our iterative approach indicates that biases remain inherent in these generations. Additionally, future research can also extend the experiment beyond our specific sample of papers and observe the impact of cross-disciplinary differences in citation practices. 

In conclusion, while LLMs can significantly aid in citation generation, they also risk amplifying existing biases and introducing new ones, potentially skewing the structuring and the dissemination of scientific knowledge. Our study underscores the necessity for developing balanced methods to interact with LLMs, incorporating diverse datasets, and implementing bias mitigation strategies. Fair prompting techniques~\citep{NEURIPS2023_8678da90}, for instance, can be employed to reduce bias, but continuous vigilance and methodological innovation are required to ensure that the integration of LLMs into academic workflows promotes accurate knowledge dissemination.

\section*{Acknowledgements}
Andres Algaba acknowledges a fellowship from the Research Foundation Flanders under Grant No.1286924N. Vincent Ginis acknowledges support from Research Foundation Flanders under Grant No.G032822N and G0K9322N.

\bibliography{biblio.bib}

\begin{thebibliography}{38}
\providecommand{\natexlab}[1]{#1}

\bibitem[{Acerbi and Stubbersfield(2023)}]{acerbi2023large}
Acerbi, A.; and Stubbersfield, J.~M. 2023.
\newblock Large language models show human-like content biases in transmission
  chain experiments.
\newblock \emph{Proceedings of the National Academy of Sciences}, 120(44):
  e2313790120.

\bibitem[{Agrawal et~al.(2024)Agrawal, Suzgun, Mackey, and
  Kalai}]{agrawal-etal-2024-language}
Agrawal, A.; Suzgun, M.; Mackey, L.; and Kalai, A. 2024.
\newblock Do Language Models Know When They{'}re Hallucinating References?
\newblock In Graham, Y.; and Purver, M., eds., \emph{Findings of the
  Association for Computational Linguistics: EACL 2024}, 912--928. St.
  Julian{'}s, Malta: Association for Computational Linguistics.

\bibitem[{Boiko, MacKnight, and Gomes(2023)}]{boiko2023emergent}
Boiko, D.~A.; MacKnight, R.; and Gomes, G. 2023.
\newblock Emergent autonomous scientific research capabilities of large
  language models.
\newblock \emph{arXiv preprint arXiv:2304.05332}.

\bibitem[{Bornmann and Daniel(2008)}]{bornmann2008citation}
Bornmann, L.; and Daniel, H.-D. 2008.
\newblock What do citation counts measure? A review of studies on citing
  behavior.
\newblock \emph{Journal of documentation}, 64(1): 45--80.

\bibitem[{Brown et~al.(2020)Brown, Mann, Ryder, Subbiah, Kaplan, Dhariwal,
  Neelakantan, Shyam, Sastry, Askell et~al.}]{brown2020language}
Brown, T.; Mann, B.; Ryder, N.; Subbiah, M.; Kaplan, J.~D.; Dhariwal, P.;
  Neelakantan, A.; Shyam, P.; Sastry, G.; Askell, A.; et~al. 2020.
\newblock Language models are few-shot learners.
\newblock \emph{Advances in neural information processing systems}, 33:
  1877--1901.

\bibitem[{Bubeck et~al.(2023)Bubeck, Chandrasekaran, Eldan, Gehrke, Horvitz,
  Kamar, Lee, Lee, Li, Lundberg et~al.}]{bubeck2023sparks}
Bubeck, S.; Chandrasekaran, V.; Eldan, R.; Gehrke, J.; Horvitz, E.; Kamar, E.;
  Lee, P.; Lee, Y.~T.; Li, Y.; Lundberg, S.; et~al. 2023.
\newblock Sparks of artificial general intelligence: Early experiments with
  gpt-4.
\newblock \emph{arXiv preprint arXiv:2303.12712}.

\bibitem[{Chen et~al.(2024)Chen, Lin, Han, and Sun}]{Chen_Lin_Han_Sun_2024}
Chen, J.; Lin, H.; Han, X.; and Sun, L. 2024.
\newblock Benchmarking Large Language Models in Retrieval-Augmented Generation.
\newblock \emph{Proceedings of the AAAI Conference on Artificial Intelligence},
  38(16): 17754--17762.

\bibitem[{Fabiano et~al.(2024)Fabiano, Gupta, Bhambra, Luu, Wong, Maaz,
  Fiedorowicz, Smith, and Solmi}]{fabiano2024optimize}
Fabiano, N.; Gupta, A.; Bhambra, N.; Luu, B.; Wong, S.; Maaz, M.; Fiedorowicz,
  J.~G.; Smith, A.~L.; and Solmi, M. 2024.
\newblock How to optimize the systematic review process using AI tools.
\newblock \emph{JCPP Advances}, e12234.

\bibitem[{Fortunato et~al.(2018)Fortunato, Bergstrom, B{\"o}rner, Evans,
  Helbing, Milojevi{\'c}, Petersen, Radicchi, Sinatra, Uzzi
  et~al.}]{fortunato2018science}
Fortunato, S.; Bergstrom, C.~T.; B{\"o}rner, K.; Evans, J.~A.; Helbing, D.;
  Milojevi{\'c}, S.; Petersen, A.~M.; Radicchi, F.; Sinatra, R.; Uzzi, B.;
  et~al. 2018.
\newblock Science of science.
\newblock \emph{Science}, 359(6379): eaao0185.

\bibitem[{Gazni and Didegah(2011)}]{gazni2011investigating}
Gazni, A.; and Didegah, F. 2011.
\newblock Investigating different types of research collaboration and citation
  impact: a case study of Harvard University's publications.
\newblock \emph{Scientometrics}, 87(2): 251--265.

\bibitem[{Jimenez et~al.(2024)Jimenez, Yang, Wettig, Yao, Pei, Press, and
  Narasimhan}]{jimenez2024swebench}
Jimenez, C.~E.; Yang, J.; Wettig, A.; Yao, S.; Pei, K.; Press, O.; and
  Narasimhan, K.~R. 2024.
\newblock {SWE}-bench: Can Language Models Resolve Real-world Github Issues?
\newblock In \emph{The Twelfth International Conference on Learning
  Representations}.

\bibitem[{Kadavath et~al.(2022)Kadavath, Conerly, Askell, Henighan, Drain,
  Perez, Schiefer, Hatfield-Dodds, DasSarma, Tran-Johnson
  et~al.}]{kadavath2022language}
Kadavath, S.; Conerly, T.; Askell, A.; Henighan, T.; Drain, D.; Perez, E.;
  Schiefer, N.; Hatfield-Dodds, Z.; DasSarma, N.; Tran-Johnson, E.; et~al.
  2022.
\newblock Language models (mostly) know what they know.
\newblock \emph{arXiv preprint arXiv:2207.05221}.

\bibitem[{Kaddour et~al.(2023)Kaddour, Harris, Mozes, Bradley, Raileanu, and
  McHardy}]{kaddour2023challenges}
Kaddour, J.; Harris, J.; Mozes, M.; Bradley, H.; Raileanu, R.; and McHardy, R.
  2023.
\newblock Challenges and applications of large language models.
\newblock \emph{arXiv preprint arXiv:2307.10169}.

\bibitem[{Kandpal et~al.(2023)Kandpal, Deng, Roberts, Wallace, and
  Raffel}]{kandpal2023large}
Kandpal, N.; Deng, H.; Roberts, A.; Wallace, E.; and Raffel, C. 2023.
\newblock Large language models struggle to learn long-tail knowledge.
\newblock In \emph{International Conference on Machine Learning}, 15696--15707.
  PMLR.

\bibitem[{Kinney et~al.(2023)Kinney, Anastasiades, Authur, Beltagy, Bragg,
  Buraczynski, Cachola, Candra, Chandrasekhar, Cohan
  et~al.}]{kinney2023semantic}
Kinney, R.; Anastasiades, C.; Authur, R.; Beltagy, I.; Bragg, J.; Buraczynski,
  A.; Cachola, I.; Candra, S.; Chandrasekhar, Y.; Cohan, A.; et~al. 2023.
\newblock The semantic scholar open data platform.
\newblock \emph{arXiv preprint arXiv:2301.10140}.

\bibitem[{Kojima et~al.(2022)Kojima, Gu, Reid, Matsuo, and
  Iwasawa}]{kojima2022large}
Kojima, T.; Gu, S.~S.; Reid, M.; Matsuo, Y.; and Iwasawa, Y. 2022.
\newblock Large language models are zero-shot reasoners.
\newblock \emph{Advances in neural information processing systems}, 35:
  22199--22213.

\bibitem[{Lawrence(2003)}]{lawrence2003politics}
Lawrence, P.~A. 2003.
\newblock The politics of publication.
\newblock \emph{Nature}, 422(6929): 259--261.

\bibitem[{Letchford, Moat, and Preis(2015)}]{letchford2015advantage}
Letchford, A.; Moat, H.~S.; and Preis, T. 2015.
\newblock The advantage of short paper titles.
\newblock \emph{Royal Society open science}, 2(8): 150266.

\bibitem[{Lewis et~al.(2020)Lewis, Perez, Piktus, Petroni, Karpukhin, Goyal,
  K{\"u}ttler, Lewis, Yih, Rockt{\"a}schel et~al.}]{lewis2020retrieval}
Lewis, P.; Perez, E.; Piktus, A.; Petroni, F.; Karpukhin, V.; Goyal, N.;
  K{\"u}ttler, H.; Lewis, M.; Yih, W.-t.; Rockt{\"a}schel, T.; et~al. 2020.
\newblock Retrieval-augmented generation for knowledge-intensive nlp tasks.
\newblock \emph{Advances in Neural Information Processing Systems}, 33:
  9459--9474.

\bibitem[{Lu et~al.(2024)Lu, Lu, Lange, Foerster, Clune, and
  Ha}]{lu2024aiscientist}
Lu, C.; Lu, C.; Lange, R.~T.; Foerster, J.; Clune, J.; and Ha, D. 2024.
\newblock The AI Scientist: Towards Fully Automated Open-Ended Scientific
  Discovery.
\newblock \emph{arXiv preprint arXiv:2408.06292}.

\bibitem[{Ma et~al.(2023)Ma, Zhang, Bian, Liu, Zhang, Zhao, Zhang, Fu, Hu, and
  Wu}]{NEURIPS2023_8678da90}
Ma, H.; Zhang, C.; Bian, Y.; Liu, L.; Zhang, Z.; Zhao, P.; Zhang, S.; Fu, H.;
  Hu, Q.; and Wu, B. 2023.
\newblock Fairness-guided Few-shot Prompting for Large Language Models.
\newblock In Oh, A.; Naumann, T.; Globerson, A.; Saenko, K.; Hardt, M.; and
  Levine, S., eds., \emph{Advances in Neural Information Processing Systems},
  volume~36, 43136--43155. Curran Associates, Inc.

\bibitem[{Manerba et~al.(2023)Manerba, Sta{\'n}czak, Guidotti, and
  Augenstein}]{manerba2023social}
Manerba, M.~M.; Sta{\'n}czak, K.; Guidotti, R.; and Augenstein, I. 2023.
\newblock Social bias probing: Fairness benchmarking for language models.
\newblock \emph{arXiv preprint arXiv:2311.09090}.

\bibitem[{Merchant et~al.(2023)Merchant, Batzner, Schoenholz, Aykol, Cheon, and
  Cubuk}]{merchant2023scaling}
Merchant, A.; Batzner, S.; Schoenholz, S.~S.; Aykol, M.; Cheon, G.; and Cubuk,
  E.~D. 2023.
\newblock Scaling deep learning for materials discovery.
\newblock \emph{Nature}, 624(7990): 80--85.

\bibitem[{Navigli, Conia, and Ross(2023)}]{navigli2023biases}
Navigli, R.; Conia, S.; and Ross, B. 2023.
\newblock Biases in large language models: origins, inventory, and discussion.
\newblock \emph{ACM Journal of Data and Information Quality}, 15(2): 1--21.

\bibitem[{Qureshi et~al.(2023)Qureshi, Shaughnessy, Gill, Robinson, Li, and
  Agai}]{qureshi2023chatgpt}
Qureshi, R.; Shaughnessy, D.; Gill, K.~A.; Robinson, K.~A.; Li, T.; and Agai,
  E. 2023.
\newblock Are ChatGPT and large language models ``the answer'' to bringing us
  closer to systematic review automation?
\newblock \emph{Systematic Reviews}, 12(1): 72.

\bibitem[{Romera-Paredes et~al.(2024)Romera-Paredes, Barekatain, Novikov,
  Balog, Kumar, Dupont, Ruiz, Ellenberg, Wang, Fawzi
  et~al.}]{romera2024mathematical}
Romera-Paredes, B.; Barekatain, M.; Novikov, A.; Balog, M.; Kumar, M.~P.;
  Dupont, E.; Ruiz, F.~J.; Ellenberg, J.~S.; Wang, P.; Fawzi, O.; et~al. 2024.
\newblock Mathematical discoveries from program search with large language
  models.
\newblock \emph{Nature}, 625(7995): 468--475.

\bibitem[{Smith(2012)}]{smith2012impact}
Smith, D.~R. 2012.
\newblock Impact factors, scientometrics and the history of citation-based
  research.
\newblock \emph{Scientometrics}, 92(2): 419--427.

\bibitem[{Srivastava et~al.(2022)Srivastava, Rastogi, Rao, Shoeb, Abid, Fisch,
  Brown, Santoro, Gupta, Garriga-Alonso et~al.}]{srivastava2022beyond}
Srivastava, A.; Rastogi, A.; Rao, A.; Shoeb, A. A.~M.; Abid, A.; Fisch, A.;
  Brown, A.~R.; Santoro, A.; Gupta, A.; Garriga-Alonso, A.; et~al. 2022.
\newblock Beyond the imitation game: Quantifying and extrapolating the
  capabilities of language models.
\newblock \emph{arXiv preprint arXiv:2206.04615}.

\bibitem[{Susnjak et~al.(2024)Susnjak, Hwang, Reyes, Barczak, McIntosh, and
  Ranathunga}]{susnjak2024automating}
Susnjak, T.; Hwang, P.; Reyes, N.~H.; Barczak, A.~L.; McIntosh, T.~R.; and
  Ranathunga, S. 2024.
\newblock Automating research synthesis with domain-specific large language
  model fine-tuning.
\newblock \emph{arXiv preprint arXiv:2404.08680}.

\bibitem[{Valenzuela-Escarcega, Ha, and
  Etzioni(2015)}]{valenzuela2015influential}
Valenzuela-Escarcega, M.~A.; Ha, V.~A.; and Etzioni, O. 2015.
\newblock Identifying Meaningful Citations.
\newblock In \emph{AAAI Workshop: Scholarly Big Data}.

\bibitem[{Walters and Wilder(2023)}]{walters2023fabrication}
Walters, W.~H.; and Wilder, E.~I. 2023.
\newblock Fabrication and errors in the bibliographic citations generated by
  ChatGPT.
\newblock \emph{Scientific Reports}, 13(1): 14045.

\bibitem[{Wang(2014)}]{wang2014unpacking}
Wang, J. 2014.
\newblock Unpacking the {M}atthew effect in citations.
\newblock \emph{Journal of Informetrics}, 8(2): 329--339.

\bibitem[{Wang et~al.(2022)Wang, Wei, Schuurmans, Le, Chi, Narang, Chowdhery,
  and Zhou}]{wang2022self}
Wang, X.; Wei, J.; Schuurmans, D.; Le, Q.; Chi, E.; Narang, S.; Chowdhery, A.;
  and Zhou, D. 2022.
\newblock Self-consistency improves chain of thought reasoning in language
  models.
\newblock \emph{arXiv preprint arXiv:2203.11171}.

\bibitem[{Wei et~al.(2022{\natexlab{a}})Wei, Tay, Bommasani, Raffel, Zoph,
  Borgeaud, Yogatama, Bosma, Zhou, Metzler et~al.}]{wei2022emergent}
Wei, J.; Tay, Y.; Bommasani, R.; Raffel, C.; Zoph, B.; Borgeaud, S.; Yogatama,
  D.; Bosma, M.; Zhou, D.; Metzler, D.; et~al. 2022{\natexlab{a}}.
\newblock Emergent abilities of large language models.
\newblock \emph{arXiv preprint arXiv:2206.07682}.

\bibitem[{Wei et~al.(2022{\natexlab{b}})Wei, Wang, Schuurmans, Bosma, Xia, Chi,
  Le, Zhou et~al.}]{wei2022chain}
Wei, J.; Wang, X.; Schuurmans, D.; Bosma, M.; Xia, F.; Chi, E.; Le, Q.~V.;
  Zhou, D.; et~al. 2022{\natexlab{b}}.
\newblock Chain-of-thought prompting elicits reasoning in large language
  models.
\newblock \emph{Advances in neural information processing systems}, 35:
  24824--24837.

\bibitem[{Yao et~al.(2024)Yao, Yu, Zhao, Shafran, Griffiths, Cao, and
  Narasimhan}]{yao2024tree}
Yao, S.; Yu, D.; Zhao, J.; Shafran, I.; Griffiths, T.; Cao, Y.; and Narasimhan,
  K. 2024.
\newblock Tree of thoughts: Deliberate problem solving with large language
  models.
\newblock \emph{Advances in Neural Information Processing Systems}, 36.

\bibitem[{Zhang et~al.(2023)Zhang, Li, Cui, Cai, Liu, Fu, Huang, Zhao, Zhang,
  Chen et~al.}]{zhang2023siren}
Zhang, Y.; Li, Y.; Cui, L.; Cai, D.; Liu, L.; Fu, T.; Huang, X.; Zhao, E.;
  Zhang, Y.; Chen, Y.; et~al. 2023.
\newblock Siren's song in the AI ocean: a survey on hallucination in large
  language models.
\newblock \emph{arXiv preprint arXiv:2309.01219}.

\bibitem[{Zheng et~al.(2023)Zheng, Koh, Ju, Nguyen, May, Webb, and
  Pan}]{zheng2023large}
Zheng, Y.; Koh, H.~Y.; Ju, J.; Nguyen, A.~T.; May, L.~T.; Webb, G.~I.; and Pan,
  S. 2023.
\newblock Large language models for scientific synthesis, inference and
  explanation.
\newblock \emph{arXiv preprint arXiv:2310.07984}.

\end{thebibliography}

\clearpage
\appendix

\section{Appendix}
We detail our data processing in Appendix \ref{sec:data} and show supplementary figures and tables.

\setcounter{secnumdepth}{2}

\section{Data}
\label{sec:data}
We describe the steps of our automated pipeline to retrieve all the necessary information for our analysis. Our data collection resulted in $166$ papers published at AAAI ($25$), NeurIPS ($72$), ICML ($38$), and ICLR ($31$) for a total of 3,066 references (see Appendix Table \ref{tab:paper_details} for a full list of included papers). The data collection pipeline uses GPT-4-0613 to postprocess parts of the data, which costs approximately 14 dollars for our experiment. Note that these steps only have to be carried out once for the data collection. However, steps 4 and 5 are also used to postprocess and enrich the information of the generated references and will need to be carried out for each run. The experiment was run on 4 November 2023 and each step was manually verified and tested. Besides using GPT-4-0613, we also ran steps 6 and 7 for GPT-4o-2024-05-13 and Claude-3-5-sonnet-20240620 on 27 July 2024.

\paragraph{Step 1. ArXiv}
We search for all papers on arXiv originally posted between 1 March 2022 and 31 October 2023 in the machine learning (cs.LG) category which refer to AAAI, NeurIPS, ICLR, or ICML in their journal reference. Note that we also verify whether we can use all these arXiv papers given their data licenses and attribute their participation in Appendix Table \ref{tab:paper_details}. We use keywords (i.e., workshop, tiny paper, 2020, 2021, track on datasets and benchmarks, and bridge) to remove papers that do not appear in the conference proceedings or earlier than 2022. We download and unzip the \emph{tar.gz} file provided by the authors to arXiv and check whether the paper exists on Semantic Scholar via title matching. We store the title, ID, and date from arXiv and Semantic Scholar. Additionally, we store all the reference titles with their corresponding ID from Semantic Scholar~\citep{kinney2023semantic}.

\paragraph{Step 2. Tex} We check whether there is a main \emph{tex} file in the unzipped paper folder by looking for a single file that contains \textbackslash begin\{document\} and \textbackslash end\{document\}. If we find a main \emph{tex} file we start the cleaning process, otherwise, we exclude the paper from our analysis. The cleaning process consists of three steps. First, we remove everything except for the author information, conference information, abstract, introduction, and references. Second, we remove figures, tables, references to sections and appendices, ... Finally, we transform all citations to numbers between square brackets. After the cleaning, we look at whether there is a \emph{bib} or \emph{bbl} file available and compile the \emph{tex} to \emph{PDF}. If neither file is available or the paper has compilation errors, we exclude the paper from our analysis (Appendix Table \ref{tab:sup_1}). Note that a \emph{bib} file allows for both \emph{PDFLatex} and \emph{bibtex} compilation, while only a \emph{bbl} file does not allow for \emph{bibtex} compilation. As a consequence papers with only a \emph{bbl} file may potentially contain papers in their reference list that are not cited in the introduction of the paper. We solve this issue in the next step.

\paragraph{Step 3. PDF} We transform the \emph{PDF} to \emph{txt} and split the main content of the paper (author information, conference information, abstract, and introduction) from the references. We then look for all in-text citations by using a regex pattern to capture numbers in between square brackets and match them with the reference list. This approach ensures that we only keep references that are cited in the introduction. We store the main content of the paper and the references cited in the introduction in separate \emph{txt} files.

\paragraph{Step 4. Postprocessing} A large number of variations and inconsistencies in the reference lists makes it difficult to structurally extract and analyze all the author information, title, publication venue, and year. We noticed that this behavior was even more outspoken in the LLM-generated references. Therefore, we examine the capabilities of GPT-4 to impose a structure on the reference list by postprocessing the data. We feed GPT-4 the reference list in \emph{txt} accompanied by the default system message: ``\emph{You are a helpful assistant}'' and the following postprocessing prompt:
\begin{tcolorbox}[mypromptbox]
\begin{center}
    Below, we share with you a list of references with their corresponding citation number between square brackets. Could you for each reference extract the authors, the number of authors, title, publication year, and publication venue? Please only return the extracted information in a markdown table with the citation number (without brackets), authors, number of authors, title, publication year, and publication venue as columns. \\
    === \\
     \textbf{\lbrack LLM generated reference list\rbrack}
\end{center}
\end{tcolorbox}
We then store the markdown table in a \emph{csv}. GPT-4 successfully structures the information and makes it more consistent, for example, by removing syllable hyphens. Sometimes a small hick-up is introduced (e.g., adding a final row with ``...''), but these are manually solved in the verification process. Note that we also prompt for the number of authors. While we can easily compute the number of authors via the meta-data from Semantic Scholar, it allows us to verify the accuracy of GPT-4 on this task as we will use it later on to postprocess the generated references where a ground truth may be unavailable.

\paragraph{Step 5. Semantic Scholar} We enrich the information from the introduction references by matching the extracted title from the \emph{csv} file in the previous step with the reference titles that we extracted from Semantic Scholar in step 1. This approach provides an extra check that GPT-4 does not change the title information in Step 4. After matching, we can use the Semantic Scholar ID to retrieve the publication venue, year, authors, citation count, influential citation count, and reference count~\citep{kinney2023semantic}. Additionally, we store the IDs of the papers to which the introduction references themselves refer.

\paragraph{Step 6. ``Vanilla'' prompting} We prompt GPT-4-0613 with the main content, which includes the author information, conference information, abstract, and introduction, accompanied by the default system message: ``\emph{You are a helpful assistant}'' and the following prompt:
\begin{tcolorbox}[mypromptbox]
\begin{center}
    Below, we share with you a written introduction to a paper and have omitted the references. Numbers between square brackets indicate citations. Can you give us a suggestion for an explicit reference associated with each number? Do not return anything except the citation number between square brackets and the corresponding reference. \\
    === \\
     \textbf{\lbrack main content\rbrack}
\end{center}
\end{tcolorbox}
We then post-process GPT-4's response to extract the title, venue, publication year, author names, and number of authors for each generated reference using the same approach as described in step 4. We repeat this ``vanilla'' approach five times for all $166$ papers.

\paragraph{Step 7. Existence check} We determine whether the generated references exist via title and author names matching with Semantic Scholar entries~\citep{kinney2023semantic}. We search Semantic Scholar for the three best matches based on the reference's title and then compute the title and author names similarity. For titles, we measure the similarity between the Semantic Scholar match and the generated reference by comparing the best matching substring. For authors, we compare them by splitting into tokens (words), removing duplicates, and then calculating the similarity based on the best partial match of the sets of tokens. In case of ``et al.,'' we only consider the first author. The similarity is computed by character-level comparison. We determined the thresholds for the title and authors scores by manually labelling $100$ matches as true or false and minimizing the false positive rate. We obtain on this sample an accuracy of $95\%$ with $5$ false positives, i.e. generated references falsely classified as non-existent.

\paragraph{Step 8. ``Iterative'' prompting} We also build on our ``vanilla'' approach, by introducing an ``iterative'' approach where we prompt GPT-4-0613 with the main content accompanied by the default system message: ``\emph{You are a helpful assistant}'' and the following prompt:
\begin{tcolorbox}[mypromptbox]
\begin{center}
    \textbf{\lbrack vanilla prompt + LLM's response \rbrack} \\
    The following references associated with these citation numbers: \\ \textbf{\lbrack numbers of non-existent generated references \rbrack} \\ do not exist. Can you replace all these non-existent references with existing ones? Keep the other references as they are. Do not return anything except the citation number between square brackets and the corresponding reference. \\
    === \\
     \textbf{\lbrack main content\rbrack}
\end{center}
\end{tcolorbox}
We again postprocess GPT-4's response using the same approach as described in steps 4, 5, and 7. The previously existing generated and the newly generated references are then merged.

\clearpage

\setcounter{figure}{0}
\renewcommand{\thefigure}{B\arabic{figure}}

\begin{figure*}[t!]%
\centering
\includegraphics[width=1.0\textwidth]{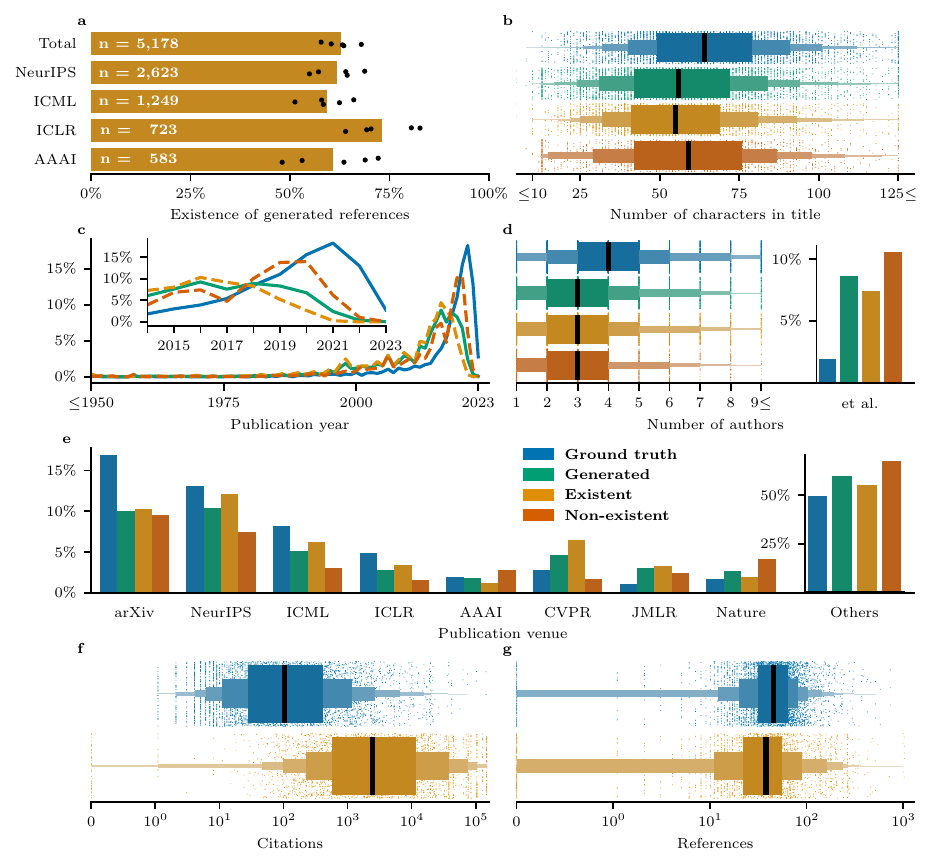}
\caption{\textbf{$\vline$ Properties of the ground truth and GPT-4 generated introduction references for the iterative strategy are consistent with the properties of the vanilla strategy.} This figure displays the properties of the ground truth ($n=5,178$, in blue) and GPT-4 generated references ($n=5,178$, in green), further subdividing the generated references into existing ($n=3,244$, in orange) and non-existent categories ($n=1,934$, in red), from the original data sources of five runs for the iterative strategy with GPT-4. Note that these are the references which are labelled ``non-existent'' in the vanilla strategy. \textbf{a}, \textbf{b}, \textbf{c}, \textbf{d}, \textbf{e}, \textbf{f} and \textbf{g,} The iterative results exhibit very similar properties to the vanilla results shown in Figure \ref{fig:main_2}.}
\label{fig:extended_1}
\end{figure*}

\begin{figure*}[t!]%
\centering
\includegraphics[width=1.0\textwidth]{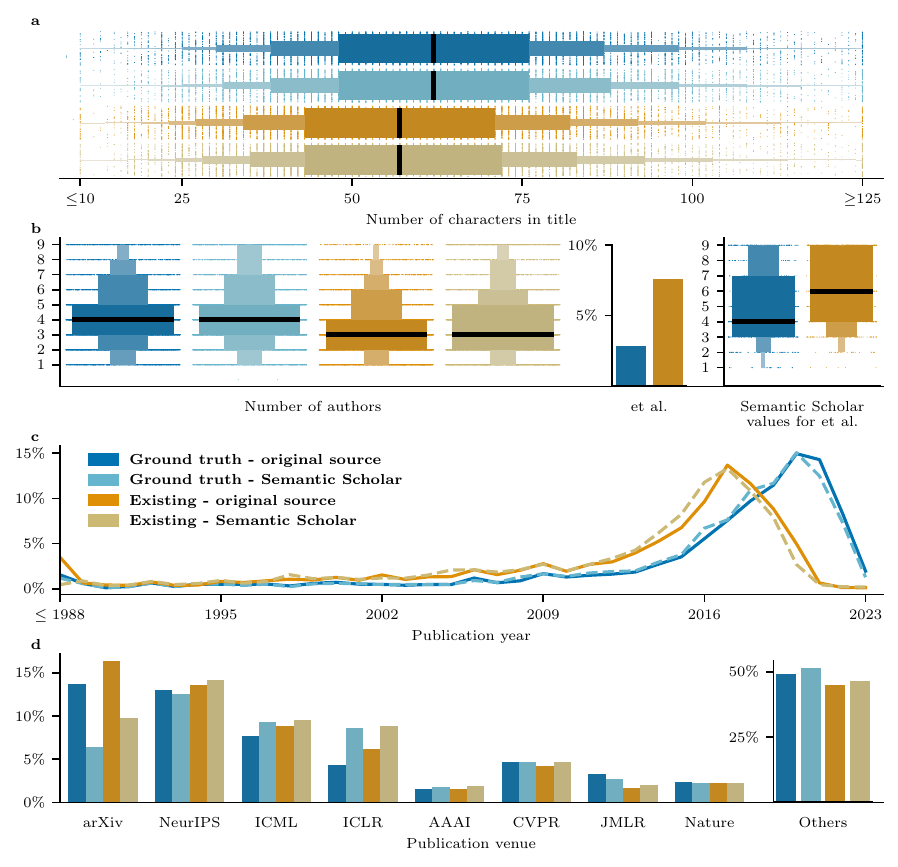}
\caption{\textbf{$\vline$ The properties of the existing GPT-4 generated references and their corresponding ground truth are consistent between the original data sources and Semantic Scholar data.} This figure compares the computation of the properties of the existing GPT-4 generated references ($n=9.376$) and their corresponding ground truth ($n=9.376$) between the data from the original sources (in dark blue and orange, as shown in Figure \ref{fig:main_2}) and from Semantic Scholar (in light blue and orange). \textbf{a,} The distributions of the number of characters in title for the existing generated references and their corresponding ground truth are very similar between the data from the original sources and Semantic Scholar. \textbf{b,} There is a discrepancy between the data from the original sources and Semantic Scholar for the number of authors in the existing generated references due to the extensive use of ``et al''. This discrepancy results in a relatively larger portion of three authors or more, but does not change the previous conclusions. \textbf{c,} The distributions over time are very similar between the data from the original sources and Semantic Scholar. \textbf{d,} There is a discrepancy between the data from the original sources and Semantic Scholar for the publication venues. The discrepancy is consistent across the existing generated references and their corresponding ground truth as both have a lower number of arXiv papers and a larger number of ICLR papers.}
\label{fig:extended_2}
\end{figure*}

\begin{figure*}[t!]%
\centering
\includegraphics[width=1.0\textwidth]{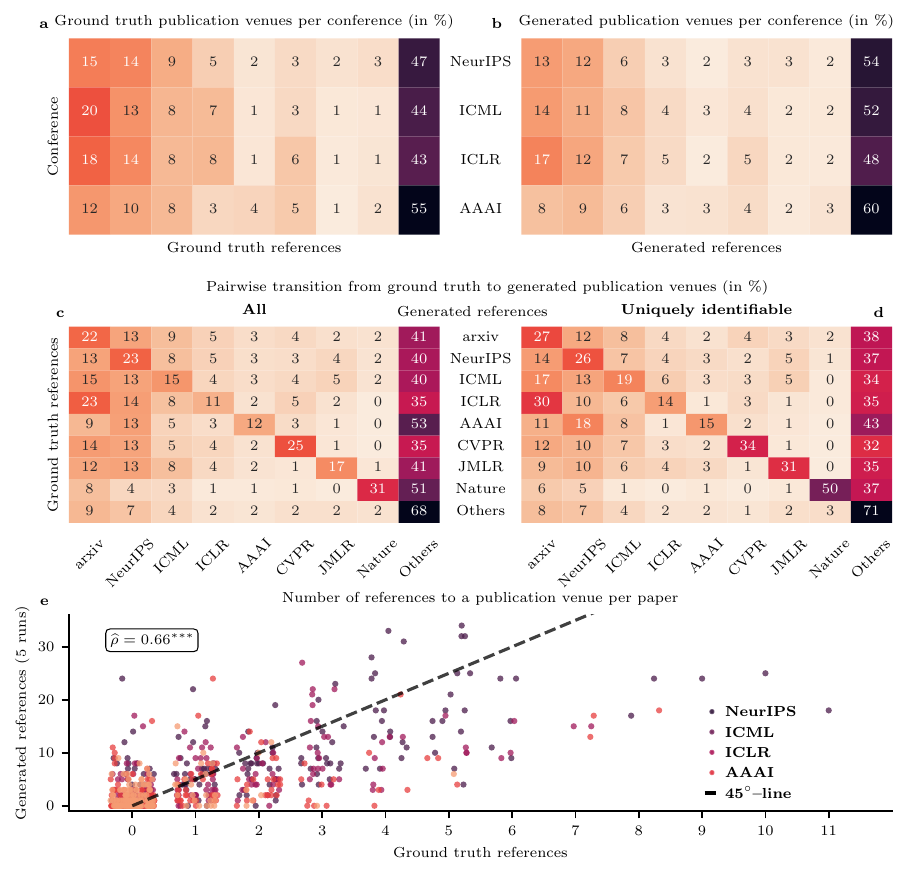}
\caption{\textbf{$\vline$ A high consistency in publication venue distributions between ground truth and GPT-4 generated references with a notable bias towards arXiv, NeurIPS, and ``Others''.} This figure displays the distributions and pairwise transition of publication venues for ground truth and GPT-4 generated references at the conference and individual paper and reference level. \textbf{a} and \textbf{b,} We observe that the distributions of publication venues for both ground truth and generated references are very similar across the various conferences, i.e., AAAI, NeurIPS, ICML, and ICLR. \textbf{c} and \textbf{d,} The pairwise transition matrix from ground truth to generated publication venues at the reference level indicates a large overall agreement, but with a strong preference in GPT-4 generated references for arXiv, NeurIPS, and ``Others'' in the case of disagreement. The preference for NeurIPS may be due to the relatively large number of NeurIPS papers in our sample and the large share of arXiv and ``Others'' points to favoring a wider array of venues which may potentially dilute the perceived relevance of key conferences. \textbf{e,} The scatter plot shows for each paper the number of ground truth references to one of the top conferences, i.e., AAAI, NeurIPS, ICML, and ICLR, and the corresponding number of generated references ($\times 5$ for five runs) which refer to the same conference.The strong pairwise correlation between the ground truth and generated references to the top conferences at the individual paper level affirms the high consistency in publication venue distributions between ground truth and GPT-4 generated references.}
\label{fig:extended_4}
\end{figure*}

\begin{figure*}[t!]%
\centering
\includegraphics[width=1.0\textwidth]{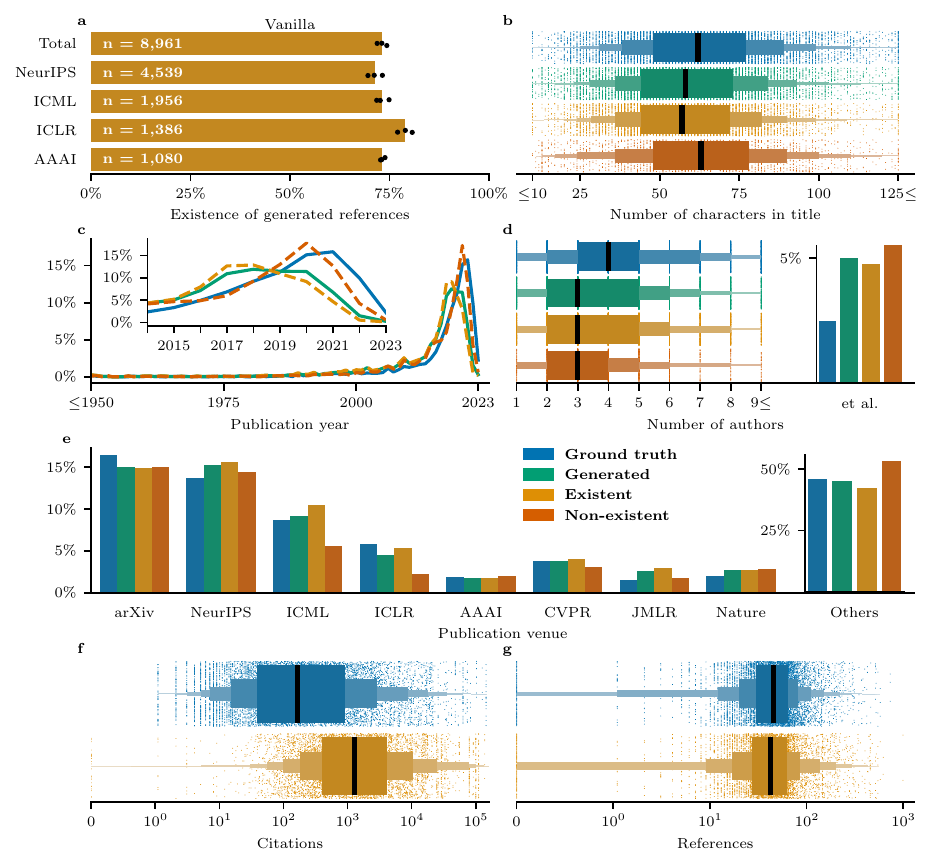}
\caption{\textbf{$\vline$ Properties of the ground truth and GPT-4o generated introduction references are consistent with the properties of GPT-4.} This figure displays the properties of the ground truth ($n=8,961$, in blue) and GPT-4o generated references ($n=8,961$, in green), further subdividing the generated references into existing ($n=6,552$, in orange) and non-existent categories ($n=2,409$, in red), from the original data sources of three runs for the vanilla strategy with GPT-4o. \textbf{a}, \textbf{b}, \textbf{c}, \textbf{d}, \textbf{e}, \textbf{f} and \textbf{g,} The GPT-4o generated references exhibit very similar properties to the GPT-4 results shown in Figure \ref{fig:main_2}, except for the existence rate which may be due to the papers now being part of the training data and the model's enhanced capabilities.}
\label{fig:extended_gpt_4o}
\end{figure*}

\begin{figure*}[t!]%
\centering
\includegraphics[width=1.0\textwidth]{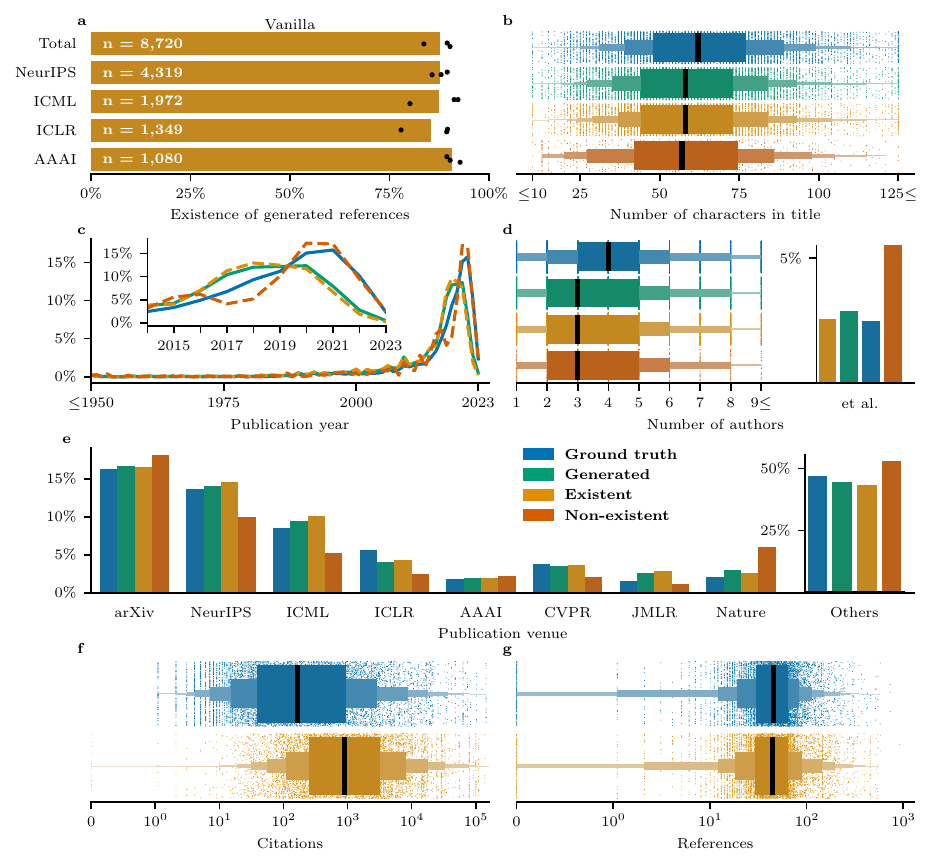}
\caption{\textbf{$\vline$ Properties of the ground truth and Claude 3.5 generated introduction references are consistent with the properties of GPT-4.} This figure displays the properties of the ground truth ($n=2,893$, in blue) and Claude 3.5 generated references ($n=2,893$, in green), further subdividing the generated references into existing ($n=2,611$, in orange) and non-existent categories ($n=282$, in red), from the original data sources of three runs for the vanilla strategy with GPT-4o. \textbf{a}, \textbf{b}, \textbf{c}, \textbf{d}, \textbf{e}, \textbf{f} and \textbf{g,} The Claude 3.5 generated references exhibit similar properties to the GPT-4 results shown in Figure \ref{fig:main_2}, except for the existence rate which may be due to the papers now being part of the training data and the model's enhanced capabilities. Additionally, Claude 3.5, on average, generates non-existent references with shorter titles and proportionally published more in arXiv, Nature, and ``others.''}
\label{fig:extended_claude_35}
\end{figure*}

\begin{figure*}[t!]%
\centering
\includegraphics[width=1.0\textwidth]{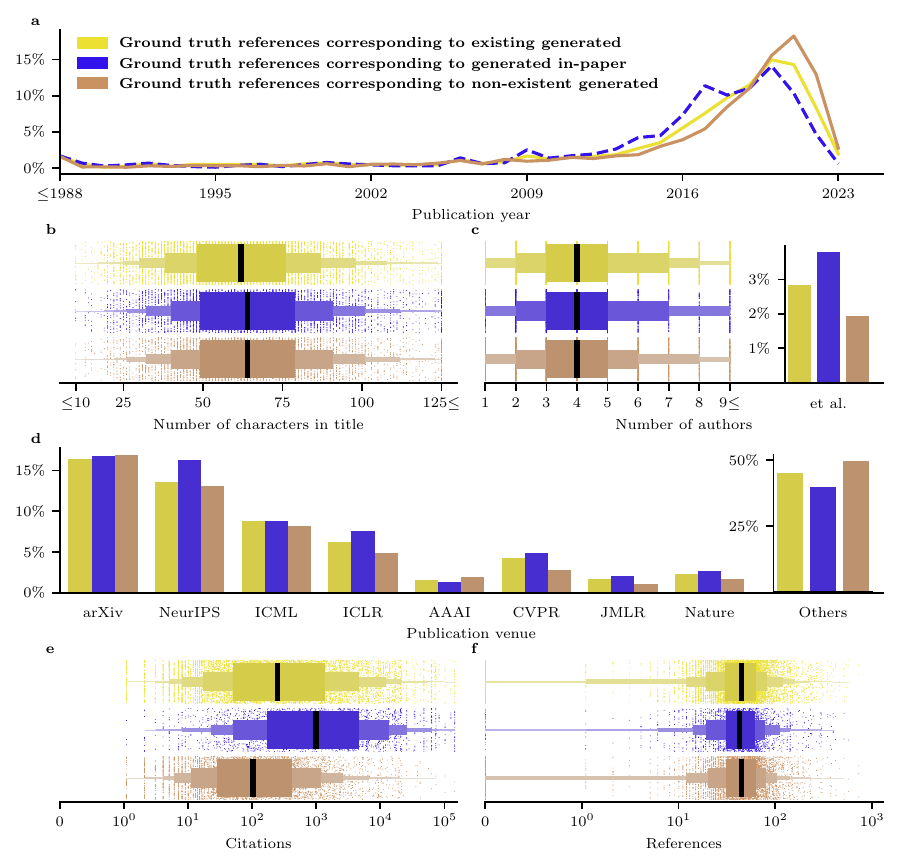}
\caption{\textbf{$\vline$ Ground truth papers which correspond to existing GPT-4 generated references that appear in the paper have substantially more citations.} This figure displays the properties of the ground truth references which correspond to existing GPT-4 generated references ($n=9,376$, in yellow), the subset of existing generated references which appear in the paper itself ($n=2,474$, in blue), and the non-existent generated references ($n=5,178$, in green), from the original data sources of five runs for the vanilla strategy with GPT-4. \textbf{a}, \textbf{b}, \textbf{c}, \textbf{d}, \textbf{e} and \textbf{f,} The ground truth papers which correspond to existing references which appear in the paper have by far the most citations, followed by the existing references, and the ground truth papers corresponding to non-existent references have the lowest numbers of citations. These findings further indicate the tendency for LLMs to more easily generate references to highly cited papers. The distributions of all other characteristics are very similar.}
\label{fig:extended_3}
\end{figure*}

\begin{figure*}[t!]%
\centering
\includegraphics[width=1.0\textwidth]{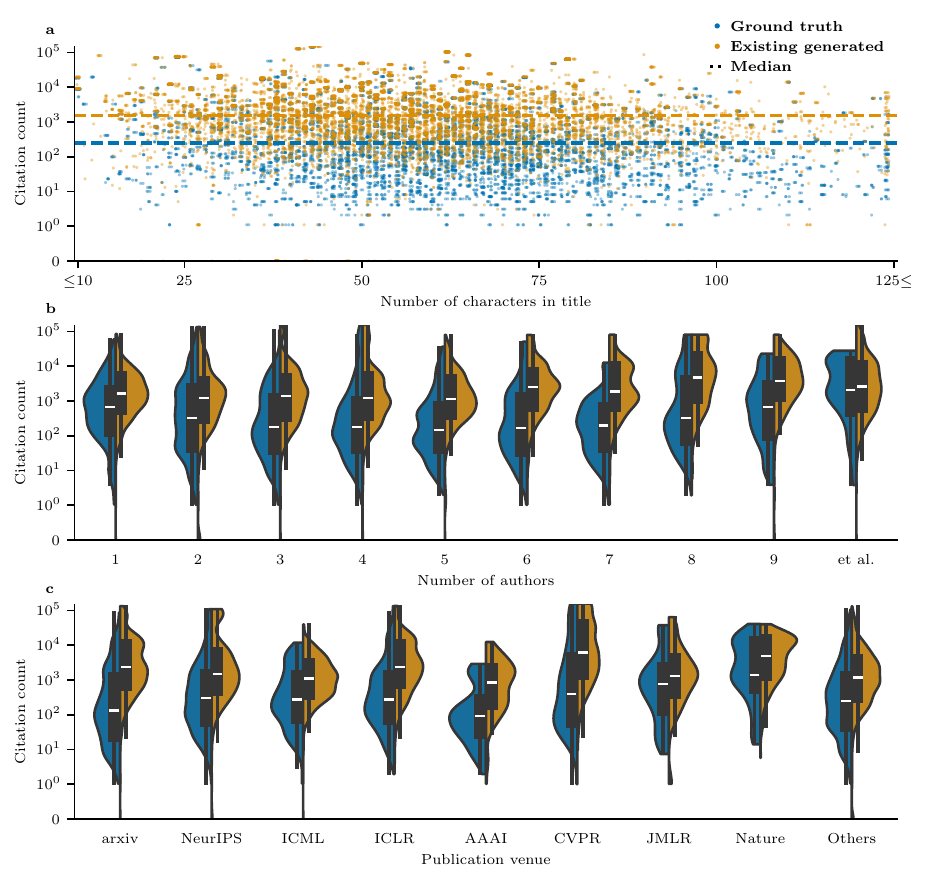}
\caption{\textbf{The citation bias in existing GPT-4 generated references is consistent across title length, number of authors, and publication venue.} This figure shows that the existing GPT-4 generated references ($n=9,376$, in orange) consistently exhibit a higher citation count compared to their corresponding ground truth ($n=9,376$, in blue) across title length, number of authors, and publication venue. \textbf{a,} The citation counts across the number of characters in title reveals that the discrepancy in number of citations between the existing generated and ground truth references is consistent over various title lengths. \textbf{b} and \textbf{c,} The distributions of citation counts per number of authors and publication venues show that the existing generated references consistently exhibit a higher citation count than their corresponding ground truth counterparts.}
\label{fig:main_4}
\end{figure*}

\begin{figure*}[t!]%
\centering
\includegraphics[width=1.0\textwidth]{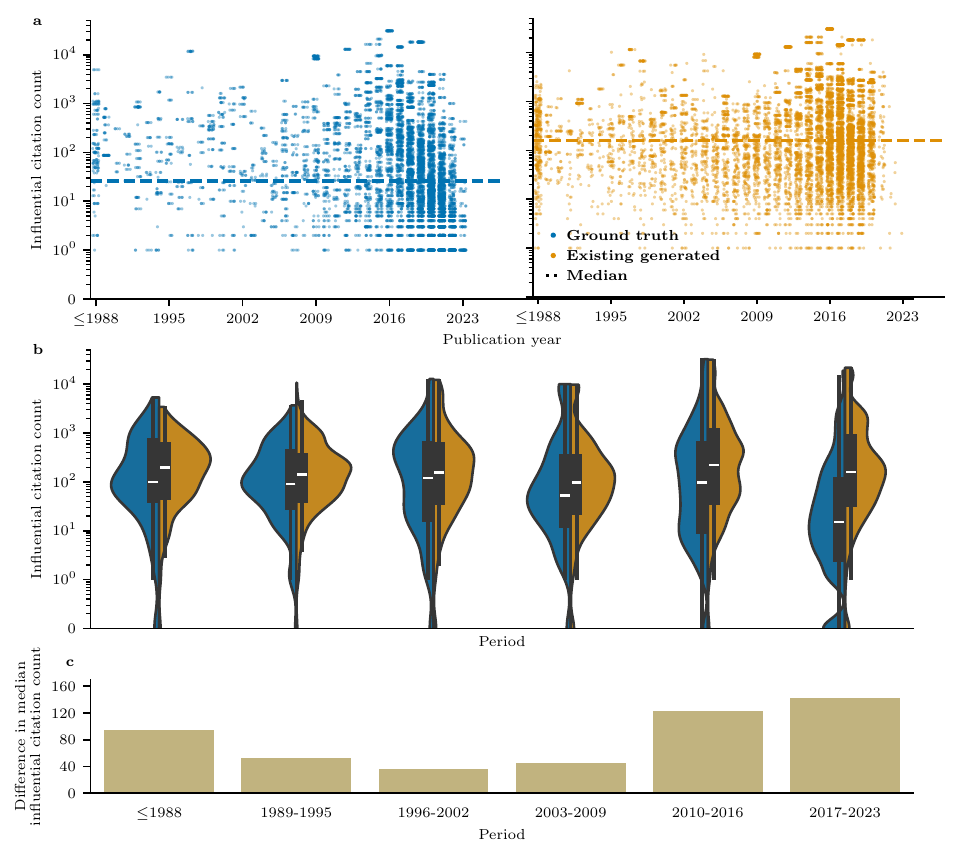}
\caption{\textbf{$\vline$ The influential citation bias in existing GPT-4 generated references is unrelated to the recency of ground truth references.} This figure shows that the existing GPT-4 generated references ($n=9,376$, in orange) consistently exhibit a higher influential citation count compared to their corresponding ground truth ($n=9,376$, in blue) across subperiods. \textbf{a}, \textbf{b} and \textbf{c,} Note that the influential citation count is retrieved from Semantic Scholar~\cite{valenzuela2015influential}.}
\label{fig:extended_5}
\end{figure*}

\begin{figure*}[t!]%
\centering
\includegraphics[width=1.0\textwidth]{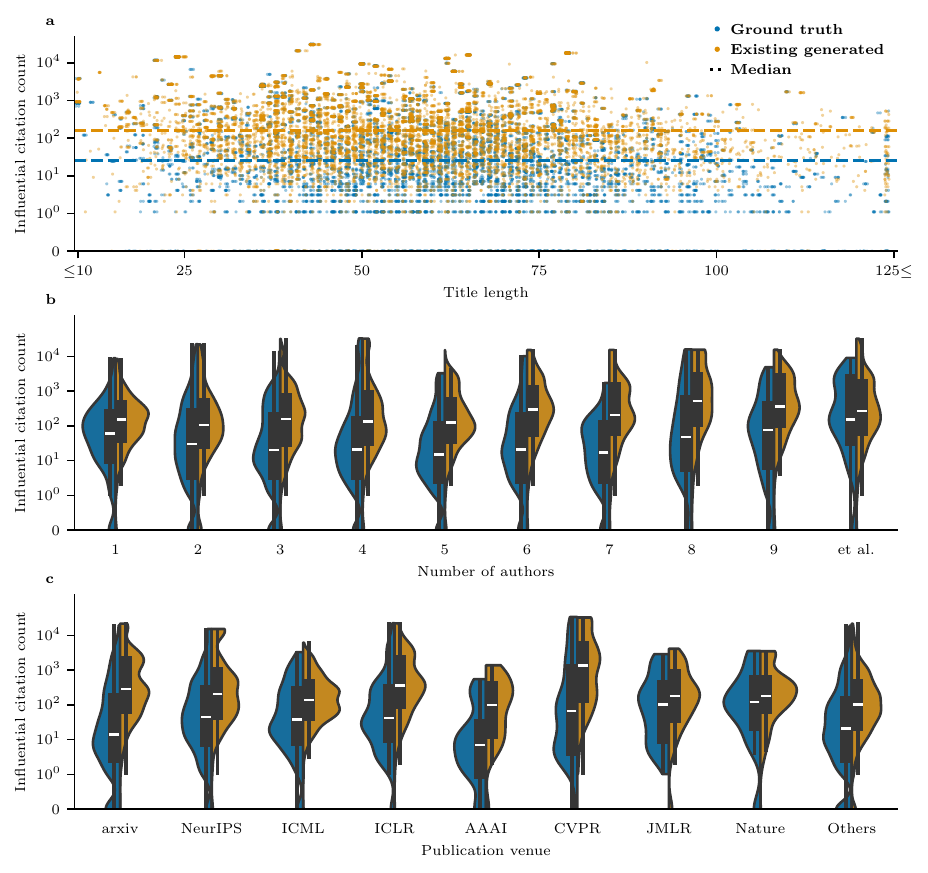}
\caption{\textbf{$\vline$ The infuential citation bias in existing GPT-4 generated references is unrelated to title length, number of authors, and publication venue.} This figure shows that the existing GPT-4 generated references ($n=9,376$, in orange) consistently exhibit a higher influential citation count compared to their corresponding ground truth ($n=9,376$, in blue) across title length, number of authors, and publication venue. \textbf{a}, \textbf{b} and \textbf{c,} Note that the influential citation count is retrieved from Semantic Scholar~\cite{valenzuela2015influential}.}
\label{fig:extended_6}
\end{figure*}

\clearpage

\setcounter{table}{0}
\renewcommand{\thetable}{C\arabic{table}}

\begin{table*}[t!]
\centering
\begin{tabular}{ccccc}
\textbf{Vanilla} & Run 1 & Run 2 & Run 3 & Run 4 \\ 
\hline
 \addlinespace
Run 2 & 17.90 & \\ 
  \addlinespace
Run 3 & 17.11 & 17.30 & \\ 
  \addlinespace
Run 4 & 17.73 & 16.69 & 16.35 \\ 
 \addlinespace
Run 5 & 18.26 & 17.78 & 17.06 & 18.44 \\
 \addlinespace
\end{tabular}
\caption{\textbf{Overlap between generated sets of references of different runs by GPT-4.} We see on average a $17\%$ overlap between different runs, which indicate that the models do not suffer from mode collapse (numbers in \% with respect to total number of references).}
\label{tab:overlap}
\end{table*}

\begin{table*}[t!]
\centering
\begin{tabular}{ccccccc}
\textbf{Vanilla} & \multicolumn{3}{c}{GPT-4o} & \multicolumn{3}{c}{Claude 3.5} \\ 
& Run 1 & Run 2 & Run 3 & Run 1 & Run 2 & Run 3 \\
\hline
 \addlinespace
Existence & 74.34 & 73.10 & 71.92 & 90.25 & 89.53 & 83.66 \\
 \addlinespace
Cited in paper & 32.16 & 33.28 & 33.71 & 42.08 & 41.50 & 39.26 \\
  \addlinespace
Cited in introduction & 24.15 & 26.23 & 25.81 & 34.23 & 33.53 & 31.85 \\ 
  \addlinespace
Pairwise Match (PM) & 10.34 & 10.62 & 10.39 & 18.04 & 17.24 & 14.74 \\
 \addlinespace
PM for uniquely & 17.49 & 19.10 & 18.28 & 29.75 & 29.25 & 25.64 \\
\end{tabular}
\caption{\textbf{Summary statistics of generated references by GPT-4o and Claude 3.5.}  (numbers in \% with respect to total number of references).}
\label{tab:extended_2}
\end{table*}

\clearpage
\onecolumn
\begin{longtable}[t!]
{p{2cm}|p{6cm}|p{7cm}}
\hline
\textbf{Conference} & \textbf{Authors} & \textbf{Title} \\
\hline
\endfirsthead

\textbf{Conference} & \textbf{Authors} & \textbf{Title} \\
\hline
\endhead

\hline
\caption{\textbf{Papers included in the analysis.}}\label{tab:paper_details} \\
\endfoot

AAAI & Jakob Weissteiner, Jakob Heiss, Julien Siems, Sven Seuken & Bayesian Optimization-based Combinatorial Assignment \\
AAAI & Gobinda Saha, Kaushik Roy & Continual Learning with Scaled Gradient Projection \\
AAAI & Ruizhe Zheng, Jun Li, Yi Wang, Tian Luo, Yuguo Yu & ScatterFormer: Locally-Invariant Scattering Transformer for Patient-Independent Multispectral Detection of Epileptiform Discharges \\
AAAI & Sahil Manchanda, Sayan Ranu & Lifelong Learning for Neural powered Mixed Integer Programming \\
AAAI & Joris Guérin, Kevin Delmas, Raul Sena Ferreira, Jérémie Guiochet & Out-Of-Distribution Detection Is Not All You Need \\
AAAI & Taha Belkhouja, Yan Yan, Janardhan Rao Doppa & Training Robust Deep Models for Time-Series Domain: Novel Algorithms and Theoretical Analysis \\
AAAI & Minsoo Kang, Suhyun Kim & GuidedMixup: An Efficient Mixup Strategy Guided by Saliency Maps \\
AAAI & Su Kim, Dongha Lee, SeongKu Kang, Seonghyeon Lee, Hwanjo Yu & Learning Topology-Specific Experts for Molecular Property Prediction \\
AAAI & Daniel Silver, Tirthak Patel, Devesh Tiwari & QUILT: Effective Multi-Class Classification on Quantum Computers Using an Ensemble of Diverse Quantum Classifiers \\
AAAI & Kevin Osanlou, Jeremy Frank, Andrei Bursuc, Tristan Cazenave, Eric Jacopin, Christophe Guettier, J. Benton & Solving Disjunctive Temporal Networks with Uncertainty under Restricted Time-Based Controllability using Tree Search and Graph Neural Networks \\
AAAI & Joar Skalse, Alessandro Abate & Misspecification in Inverse Reinforcement Learning \\
AAAI & Edward Ayers, Jonathan Sadeghi, John Redford, Romain Mueller, Puneet K. Dokania & Query-based Hard-Image Retrieval for Object Detection at Test Time \\
AAAI & Shubham Gupta, Sahil Manchanda, Srikanta Bedathur, Sayan Ranu & TIGGER: Scalable Generative Modelling for Temporal Interaction Graphs \\
AAAI & Fanchen Bu, Dong Eui Chang & Feedback Gradient Descent: Efficient and Stable Optimization with Orthogonality for DNNs \\
AAAI & Shota Saito & Hypergraph Modeling via Spectral Embedding Connection: Hypergraph Cut, Weighted Kernel $k$-means, and Heat Kernel \\
AAAI & Haoran Luo, Haihong E, Ling Tan, Gengxian Zhou, Tianyu Yao, Kaiyang Wan & DHGE: Dual-View Hyper-Relational Knowledge Graph Embedding for Link Prediction and Entity Typing \\
AAAI & Yujin Kim, Dogyun Park, Dohee Kim, Suhyun Kim & NaturalInversion: Data-Free Image Synthesis Improving Real-World Consistency \\
AAAI & Tairan He, Weiye Zhao, Changliu Liu & AutoCost: Evolving Intrinsic Cost for Zero-violation Reinforcement Learning \\
AAAI & Shijie Liu, Andrew C. Cullen, Paul Montague, Sarah M. Erfani, Benjamin I. P. Rubinstein & Enhancing the Antidote: Improved Pointwise Certifications against Poisoning Attacks \\
AAAI & Fan Zhou, Chen Pan, Lintao Ma, Yu Liu, Shiyu Wang, James Zhang, Xinxin Zhu, Xuanwei Hu, Yunhua Hu, Yangfei Zheng, Lei Lei, Yun Hu & SLOTH: Structured Learning and Task-based Optimization for Time Series Forecasting on Hierarchies \\
AAAI & Christopher W. F. Parsonson, Alexandre Laterre, Thomas D. Barrett & Reinforcement Learning for Branch-and-Bound Optimisation using Retrospective Trajectories \\
AAAI & Sourya Basu, Prasanna Sattigeri, Karthikeyan Natesan Ramamurthy, Vijil Chenthamarakshan, Kush R. Varshney, Lav R. Varshney, Payel Das & Equi-Tuning: Group Equivariant Fine-Tuning of Pretrained Models \\
AAAI & Harry Rubin-Falcone, Joyce Lee, Jenna Wiens & Forecasting with Sparse but Informative Variables: A Case Study in Predicting Blood Glucose \\
AAAI & Pierre Le Pelletier de Woillemont, Rémi Labory, Vincent Corruble & Automated Play-Testing Through RL Based Human-Like Play-Styles Generation \\
AAAI & Kai Klede, Leo Schwinn, Dario Zanca, Björn Eskofier & FastAMI -- a Monte Carlo Approach to the Adjustment for Chance in Clustering Comparison Metrics \\
NeurIPS & Dhananjay Bhaskar, Kincaid MacDonald, Oluwadamilola Fasina, Dawson Thomas, Bastian Rieck, Ian Adelstein, Smita Krishnaswamy & Diffusion Curvature for Estimating Local Curvature in High Dimensional Data \\
NeurIPS & Shiro Takagi & On the Effect of Pre-training for Transformer in Different Modality on Offline Reinforcement Learning \\
NeurIPS & Yue Yu, Yuchen Zhuang, Jieyu Zhang, Yu Meng, Alexander Ratner, Ranjay Krishna, Jiaming Shen, Chao Zhang & Large Language Model as Attributed Training Data Generator: A Tale of Diversity and Bias \\
NeurIPS & Lingfeng Sun, Haichao Zhang, Wei Xu, Masayoshi Tomizuka & PaCo: Parameter-Compositional Multi-Task Reinforcement Learning \\
NeurIPS & Yang Yue, Rui Lu, Bingyi Kang, Shiji Song, Gao Huang & Understanding, Predicting and Better Resolving Q-Value Divergence in Offline-RL \\
NeurIPS & Jiaqi Leng, Yuxiang Peng, Yi-Ling Qiao, Ming Lin, Xiaodi Wu & Differentiable Analog Quantum Computing for Optimization and Control \\
NeurIPS & Kyriakos Flouris, Ender Konukoglu & Canonical normalizing flows for manifold learning \\
NeurIPS & Yuchen Bai, Jean-Baptiste Durand, Florence Forbes, Grégoire Vincent & Semantic segmentation of sparse irregular point clouds for leaf wood discrimination \\
NeurIPS & Lorenzo Giambagli, Lorenzo Buffoni, Lorenzo Chicchi, Duccio Fanelli & How a student becomes a teacher: learning and forgetting through Spectral methods \\
NeurIPS & Hanbyul Lee, Qifan Song, Jean Honorio & Support Recovery in Sparse PCA with Incomplete Data \\
NeurIPS & Zhang-Wei Hong, Aviral Kumar, Sathwik Karnik, Abhishek Bhandwaldar, Akash Srivastava, Joni Pajarinen, Romain Laroche, Abhishek Gupta, Pulkit Agrawal & Beyond Uniform Sampling: Offline Reinforcement Learning with Imbalanced Datasets \\
NeurIPS & Xiang Zhang, Ziyuan Zhao, Theodoros Tsiligkaridis, Marinka Zitnik & Self-Supervised Contrastive Pre-Training For Time Series via Time-Frequency Consistency \\
NeurIPS & Antonin Schrab, Ilmun Kim, Benjamin Guedj, Arthur Gretton & Efficient Aggregated Kernel Tests using Incomplete $U$-statistics \\
NeurIPS & Wanyun Cui, Xingran Chen & Instance-based Learning for Knowledge Base Completion \\
NeurIPS & Aurelien Lucchi, Frank Proske, Antonio Orvieto, Francis Bach, Hans Kersting & On the Theoretical Properties of Noise Correlation in Stochastic Optimization \\
NeurIPS & Minsik Cho, Saurabh Adya, Devang Naik & PDP: Parameter-free Differentiable Pruning is All You Need \\
NeurIPS & Guangxi Li, Ruilin Ye, Xuanqiang Zhao, Xin Wang & Concentration of Data Encoding in Parameterized Quantum Circuits \\
NeurIPS & Xinrui Wang, Wenhai Wan, Chuanxin Geng, Shaoyuan LI, Songcan Chen & Beyond Myopia: Learning from Positive and Unlabeled Data through Holistic Predictive Trends \\
NeurIPS & Zihan Liu, Yun Luo, Lirong Wu, Zicheng Liu, Stan Z. Li & Towards Reasonable Budget Allocation in Untargeted Graph Structure Attacks via Gradient Debias \\
NeurIPS & Dingfan Chen, Raouf Kerkouche, Mario Fritz & Private Set Generation with Discriminative Information \\
NeurIPS & Zhan Yu, Hongshun Yao, Mujin Li, Xin Wang & Power and limitations of single-qubit native quantum neural networks \\
NeurIPS & Ibrahim Alabdulmohsin, Xiaohua Zhai, Alexander Kolesnikov, Lucas Beyer & Getting ViT in Shape: Scaling Laws for Compute-Optimal Model Design \\
NeurIPS & Manzil Zaheer, Kenneth Marino, Will Grathwohl, John Schultz, Wendy Shang, Sheila Babayan, Arun Ahuja, Ishita Dasgupta, Christine Kaeser-Chen, Rob Fergus & Learning to Navigate Wikipedia by Taking Random Walks \\
NeurIPS & Dohyun Kwon, Ying Fan, Kangwook Lee & Score-based Generative Modeling Secretly Minimizes the Wasserstein Distance \\
NeurIPS & Zhaoqi Li, Lillian Ratliff, Houssam Nassif, Kevin Jamieson, Lalit Jain & Instance-optimal PAC Algorithms for Contextual Bandits \\
NeurIPS & Masaki Adachi, Satoshi Hayakawa, Martin Jørgensen, Harald Oberhauser, Michael A. Osborne & Fast Bayesian Inference with Batch Bayesian Quadrature via Kernel Recombination \\
NeurIPS & Zhiqin Yang, Yonggang Zhang, Yu Zheng, Xinmei Tian, Hao Peng, Tongliang Liu, Bo Han & FedFed: Feature Distillation against Data Heterogeneity in Federated Learning \\
NeurIPS & Daniel Vial, Sujay Sanghavi, Sanjay Shakkottai, R. Srikant & Minimax Regret for Cascading Bandits \\
NeurIPS & Fabian Zaiser, Andrzej S. Murawski, Luke Ong & Exact Bayesian Inference on Discrete Models via Probability Generating Functions: A Probabilistic Programming Approach \\
NeurIPS & Cheng Chi, Amine Mohamed Aboussalah, Elias B. Khalil, Juyoung Wang, Zoha Sherkat-Masoumi & A Deep Reinforcement Learning Framework For Column Generation \\
NeurIPS & Mathieu Molina, Patrick Loiseau & Bounding and Approximating Intersectional Fairness through Marginal Fairness \\
NeurIPS & Shuai Zhang, Hongkang Li, Meng Wang, Miao Liu, Pin-Yu Chen, Songtao Lu, Sijia Liu, Keerthiram Murugesan, Subhajit Chaudhury & On the Convergence and Sample Complexity Analysis of Deep Q-Networks with $\varepsilon$-Greedy Exploration \\
NeurIPS & Changlong Wu, Mohsen Heidari, Ananth Grama, Wojciech Szpankowski & Precise Regret Bounds for Log-loss via a Truncated Bayesian Algorithm \\
NeurIPS & Ching-Yao Chuang, Stefanie Jegelka & Tree Mover's Distance: Bridging Graph Metrics and Stability of Graph Neural Networks \\
NeurIPS & Felix Biggs, Antonin Schrab, Arthur Gretton & MMD-FUSE: Learning and Combining Kernels for Two-Sample Testing Without Data Splitting \\
NeurIPS & Thomas Fel, Victor Boutin, Mazda Moayeri, Rémi Cadène, Louis Bethune, Léo andéol, Mathieu Chalvidal, Thomas Serre & A Holistic Approach to Unifying Automatic Concept Extraction and Concept Importance Estimation \\
NeurIPS & Manel Baradad, Chun-Fu Chen, Jonas Wulff, Tongzhou Wang, Rogerio Feris, Antonio Torralba, Phillip Isola & Procedural Image Programs for Representation Learning \\
NeurIPS & Yang Ni & Bivariate Causal Discovery for Categorical Data via Classification with Optimal Label Permutation \\
NeurIPS & Gauthier Guinet, Saurabh Amin, Patrick Jaillet & Effective Dimension in Bandit Problems under Censorship \\
NeurIPS & Kyungmin Lee, Jinwoo Shin & RenyiCL: Contrastive Representation Learning with Skew Renyi Divergence \\
NeurIPS & Yihe Wang, Yu Han, Haishuai Wang, Xiang Zhang & Contrast Everything: A Hierarchical Contrastive Framework for Medical Time-Series \\
NeurIPS & Artyom Sorokin, Nazar Buzun, Leonid Pugachev, Mikhail Burtsev & Explain My Surprise: Learning Efficient Long-Term Memory by Predicting Uncertain Outcomes \\
NeurIPS & Yipeng Kang, Tonghan Wang, Xiaoran Wu, Qianlan Yang, Chongjie Zhang & Non-Linear Coordination Graphs \\
NeurIPS & Niv Giladi, Shahar Gottlieb, Moran Shkolnik, Asaf Karnieli, Ron Banner, Elad Hoffer, Kfir Yehuda Levy, Daniel Soudry & DropCompute: simple and more robust distributed synchronous training via compute variance reduction \\
NeurIPS & Jack Richter-Powell, Yaron Lipman, Ricky T. Q. Chen & Neural Conservation Laws: A Divergence-Free Perspective \\
NeurIPS & Peide Huang, Mengdi Xu, Jiacheng Zhu, Laixi Shi, Fei Fang, Ding Zhao & Curriculum Reinforcement Learning using Optimal Transport via Gradual Domain Adaptation \\
NeurIPS & Mark D. McDonnell, Dong Gong, Amin Parveneh, Ehsan Abbasnejad, Anton van den Hengel & RanPAC: Random Projections and Pre-trained Models for Continual Learning \\
NeurIPS & Haoyuan Sun, Kwangjun Ahn, Christos Thrampoulidis, Navid Azizan & Mirror Descent Maximizes Generalized Margin and Can Be Implemented Efficiently \\
NeurIPS & Rui M. Castro, Fredrik Hellström, Tim van Erven & Adaptive Selective Sampling for Online Prediction with Experts \\
NeurIPS & Tonghan Wang, Paul Dütting, Dmitry Ivanov, Inbal Talgam-Cohen, David C. Parkes & Deep Contract Design via Discontinuous Networks \\
NeurIPS & Sourya Basu, Pulkit Katdare, Prasanna Sattigeri, Vijil Chenthamarakshan, Katherine Driggs-Campbell, Payel Das, Lav R. Varshney & Efficient Equivariant Transfer Learning from Pretrained Models \\
NeurIPS & Qianyi Li, Haim Sompolinsky & Globally Gated Deep Linear Networks \\
NeurIPS & Jonatha Anselmi, Bruno Gaujal, Louis-Sébastien Rebuffi & Reinforcement Learning in a Birth and Death Process: Breaking the Dependence on the State Space \\
NeurIPS & Zhanpeng Zhou, Yongyi Yang, Xiaojiang Yang, Junchi Yan, Wei Hu & Going Beyond Linear Mode Connectivity: The Layerwise Linear Feature Connectivity \\
NeurIPS & Jinyu Cai, Jicong Fan & Perturbation Learning Based Anomaly Detection \\
NeurIPS & Dan Zhao & Combining Explicit and Implicit Regularization for Efficient Learning in Deep Networks \\
NeurIPS & Leonard Papenmeier, Luigi Nardi, Matthias Poloczek & Increasing the Scope as You Learn: Adaptive Bayesian Optimization in Nested Subspaces \\
NeurIPS & Yang Song, Qiyu Kang, Sijie Wang, Zhao Kai, Wee Peng Tay & On the Robustness of Graph Neural Diffusion to Topology Perturbations \\
NeurIPS & Ibrahim Alabdulmohsin, Behnam Neyshabur, Xiaohua Zhai & Revisiting Neural Scaling Laws in Language and Vision \\
NeurIPS & Salva Rühling Cachay, Bo Zhao, Hailey Joren, Rose Yu & DYffusion: A Dynamics-informed Diffusion Model for Spatiotemporal Forecasting \\
NeurIPS & Indradyumna Roy, Soumen Chakrabarti, Abir De & Maximum Common Subgraph Guided Graph Retrieval: Late and Early Interaction Networks \\
NeurIPS & Divin Yan, Gengchen Wei, Chen Yang, Shengzhong Zhang, Zengfeng Huang & Rethinking Semi-Supervised Imbalanced Node Classification from Bias-Variance Decomposition \\
NeurIPS & Zhiying Lu, Hongtao Xie, Chuanbin Liu, Yongdong Zhang & Bridging the Gap Between Vision Transformers and Convolutional Neural Networks on Small Datasets \\
NeurIPS & Rémi Leluc, François Portier, Johan Segers, Aigerim Zhuman & A Quadrature Rule combining Control Variates and Adaptive Importance Sampling \\
NeurIPS & Kwangjun Ahn, Xiang Cheng, Hadi Daneshmand, Suvrit Sra & Transformers learn to implement preconditioned gradient descent for in-context learning \\
NeurIPS & Annie S. Chen, Archit Sharma, Sergey Levine, Chelsea Finn & You Only Live Once: Single-Life Reinforcement Learning \\
NeurIPS & Sen Lin, Daouda Sow, Kaiyi Ji, Yingbin Liang, Ness Shroff & Non-Convex Bilevel Optimization with Time-Varying Objective Functions \\
NeurIPS & Carl Hvarfner, Erik Hellsten, Frank Hutter, Luigi Nardi & Self-Correcting Bayesian Optimization through Bayesian Active Learning \\
NeurIPS & Abir De, Soumen Chakrabarti & Neural Estimation of Submodular Functions with Applications to Differentiable Subset Selection \\
NeurIPS & Ximing Lu, Sean Welleck, Jack Hessel, Liwei Jiang, Lianhui Qin, Peter West, Prithviraj Ammanabrolu, Yejin Choi & Quark: Controllable Text Generation with Reinforced Unlearning \\
NeurIPS & Weirui Ye, Pieter Abbeel, Yang Gao & Spending Thinking Time Wisely: Accelerating MCTS with Virtual Expansions \\
NeurIPS & Axel Levy, Gordon Wetzstein, Julien Martel, Frederic Poitevin, Ellen D. Zhong & Amortized Inference for Heterogeneous Reconstruction in Cryo-EM \\
ICLR & Tim Pearce, Tabish Rashid, Anssi Kanervisto, Dave Bignell, Mingfei Sun, Raluca Georgescu, Sergio Valcarcel Macua, Shan Zheng Tan, Ida Momennejad, Katja Hofmann, Sam Devlin & Imitating Human Behaviour with Diffusion Models \\
ICLR & Yi Ren, Shangmin Guo, Wonho Bae, Danica J. Sutherland & How to prepare your task head for finetuning \\
ICLR & Kieran A. Murphy, Dani S. Bassett & Interpretability with full complexity by constraining feature information \\
ICLR & Julius Adebayo, Michael Muelly, Hal Abelson, Been Kim & Post hoc Explanations may be Ineffective for Detecting Unknown Spurious Correlation \\
ICLR & Roman Levin, Valeriia Cherepanova, Avi Schwarzschild, Arpit Bansal, C. Bayan Bruss, Tom Goldstein, Andrew Gordon Wilson, Micah Goldblum & Transfer Learning with Deep Tabular Models \\
ICLR & Aviv A. Rosenberg, Sanketh Vedula, Yaniv Romano, Alex M. Bronstein & Fast Nonlinear Vector Quantile Regression \\
ICLR & Edward De Brouwer, Rahul G. Krishnan & Anamnesic Neural Differential Equations with Orthogonal Polynomial Projections \\
ICLR & Ilya Trofimov, Daniil Cherniavskii, Eduard Tulchinskii, Nikita Balabin, Evgeny Burnaev, Serguei Barannikov & Learning Topology-Preserving Data Representations \\
ICLR & Trenton Bricken, Xander Davies, Deepak Singh, Dmitry Krotov, Gabriel Kreiman & Sparse Distributed Memory is a Continual Learner \\
ICLR & Steeven Janny, Aurélien Béneteau, Madiha Nadri, Julie Digne, Nicolas Thome, Christian Wolf & Eagle: Large-Scale Learning of Turbulent Fluid Dynamics with Mesh Transformers \\
ICLR & Clement Vignac, Igor Krawczuk, Antoine Siraudin, Bohan Wang, Volkan Cevher, Pascal Frossard & DiGress: Discrete Denoising diffusion for graph generation \\
ICLR & Xinting Hu, Yulei Niu, Chunyan Miao, Xian-Sheng Hua, Hanwang Zhang & On Non-Random Missing Labels in Semi-Supervised Learning \\
ICLR & Jiefeng Chen, Timothy Nguyen, Dilan Gorur, Arslan Chaudhry & Is forgetting less a good inductive bias for forward transfer? \\
ICLR & Matthew J. Tilley, Michelle Miller, David J. Freedman & Artificial Neuronal Ensembles with Learned Context Dependent Gating \\
ICLR & Zhang-Wei Hong, Tao Chen, Yen-Chen Lin, Joni Pajarinen, Pulkit Agrawal & Topological Experience Replay \\
ICLR & Lingkai Kong, Yuqing Wang, Molei Tao & Momentum Stiefel Optimizer, with Applications to Suitably-Orthogonal Attention, and Optimal Transport \\
ICLR & Zhang-Wei Hong, Pulkit Agrawal, Rémi Tachet des Combes, Romain Laroche & Harnessing Mixed Offline Reinforcement Learning Datasets via Trajectory Weighting \\
ICLR & Meng Cao, Mehdi Fatemi, Jackie Chi Kit Cheung, Samira Shabanian & Systematic Rectification of Language Models via Dead-end Analysis \\
ICLR & Wentao Zhang, Yexin Wang, Zhenbang You, Meng Cao, Ping Huang, Jiulong Shan, Zhi Yang, Bin Cui & Information Gain Propagation: a new way to Graph Active Learning with Soft Labels \\
ICLR & Alexandre Perez-Lebel, Marine Le Morvan, Gaël Varoquaux & Beyond calibration: estimating the grouping loss of modern neural networks \\
ICLR & Jianwen Xie, Yaxuan Zhu, Jun Li, Ping Li & A Tale of Two Flows: Cooperative Learning of Langevin Flow and Normalizing Flow Toward Energy-Based Model \\
ICLR & Amrith Setlur, Don Dennis, Benjamin Eysenbach, Aditi Raghunathan, Chelsea Finn, Virginia Smith, Sergey Levine & Bitrate-Constrained DRO: Beyond Worst Case Robustness To Unknown Group Shifts \\
ICLR & Tim Z. Xiao, Robert Bamler & Trading Information between Latents in Hierarchical Variational Autoencoders \\
ICLR & Zixuan Ke, Yijia Shao, Haowei Lin, Tatsuya Konishi, Gyuhak Kim, Bing Liu & Continual Pre-training of Language Models \\
ICLR & Zhang-Wei Hong, Ge Yang, Pulkit Agrawal & Bilinear value networks \\
ICLR & Hanrong Ye, Dan Xu & Joint 2D-3D Multi-Task Learning on Cityscapes-3D: 3D Detection, Segmentation, and Depth Estimation \\
ICLR & Mohit Vaishnav, Thomas Serre & GAMR: A Guided Attention Model for (visual) Reasoning \\
ICLR & Noam Levi, Itay M. Bloch, Marat Freytsis, Tomer Volansky & Noise Injection Node Regularization for Robust Learning \\
ICLR & Paul F. Jaeger, Carsten T. Lüth, Lukas Klein, Till J. Bungert & A Call to Reflect on Evaluation Practices for Failure Detection in Image Classification \\
ICLR & Thomas M. Sutter, Laura Manduchi, Alain Ryser, Julia E. Vogt & Learning Group Importance using the Differentiable Hypergeometric Distribution \\
ICLR & AmirEhsan Khorashadizadeh, Anadi Chaman, Valentin Debarnot, Ivan Dokmanić & FunkNN: Neural Interpolation for Functional Generation \\
ICML & Ramki Gummadi, Saurabh Kumar, Junfeng Wen, Dale Schuurmans & A Parametric Class of Approximate Gradient Updates for Policy Optimization \\
ICML & Joshua P. Zitovsky, Daniel de Marchi, Rishabh Agarwal, Michael R. Kosorok & Revisiting Bellman Errors for Offline Model Selection \\
ICML & Jiayin Jin, Zeru Zhang, Yang Zhou, Lingfei Wu & Input-agnostic Certified Group Fairness via Gaussian Parameter Smoothing \\
ICML & Ching-Yao Chuang, Stefanie Jegelka, David Alvarez-Melis & InfoOT: Information Maximizing Optimal Transport \\
ICML & Ilgee Hong, Sen Na, Michael W. Mahoney, Mladen Kolar & Constrained Optimization via Exact Augmented Lagrangian and Randomized Iterative Sketching \\
ICML & Matthew Fahrbach, Adel Javanmard, Vahab Mirrokni, Pratik Worah & Learning Rate Schedules in the Presence of Distribution Shift \\
ICML & Samuele Marro, Michele Lombardi & Computational Asymmetries in Robust Classification \\
ICML & Nicolas Chopin, Andras Fulop, Jeremy Heng, Alexandre H. Thiery & Computational Doob's h-transforms for Online Filtering of Discretely Observed Diffusions \\
ICML & Wentao Zhang, Zeang Sheng, Mingyu Yang, Yang Li, Yu Shen, Zhi Yang, Bin Cui & NAFS: A Simple yet Tough-to-beat Baseline for Graph Representation Learning \\
ICML & Disha Shrivastava, Hugo Larochelle, Daniel Tarlow & Repository-Level Prompt Generation for Large Language Models of Code \\
ICML & Chenlu Ye, Wei Xiong, Quanquan Gu, Tong Zhang & Corruption-Robust Algorithms with Uncertainty Weighting for Nonlinear Contextual Bandits and Markov Decision Processes \\
ICML & Anas Barakat, Ilyas Fatkhullin, Niao He & Reinforcement Learning with General Utilities: Simpler Variance Reduction and Large State-Action Space \\
ICML & Alberto Maria Metelli, Francesco Trovò, Matteo Pirola, Marcello Restelli & Stochastic Rising Bandits \\
ICML & Idan Shenfeld, Zhang-Wei Hong, Aviv Tamar, Pulkit Agrawal & TGRL: An Algorithm for Teacher Guided Reinforcement Learning \\
ICML & Benjamin Dupuis, George Deligiannidis, Umut Şimşekli & Generalization Bounds with Data-dependent Fractal Dimensions \\
ICML & Wanrong Zhang, Ruqi Zhang & DP-Fast MH: Private, Fast, and Accurate Metropolis-Hastings for Large-Scale Bayesian Inference \\
ICML & Zixuan Ni, Longhui Wei, Siliang Tang, Yueting Zhuang, Qi Tian & Continual Vision-Language Representation Learning with Off-Diagonal Information \\
ICML & Manuel Nonnenmacher, Lukas Oldenburg, Ingo Steinwart, David Reeb & Utilizing Expert Features for Contrastive Learning of Time-Series Representations \\
ICML & Shih-Yang Liu, Zechun Liu, Kwang-Ting Cheng & Oscillation-free Quantization for Low-bit Vision Transformers \\
ICML & Siqi Liu, Marc Lanctot, Luke Marris, Nicolas Heess & Simplex Neural Population Learning: Any-Mixture Bayes-Optimality in Symmetric Zero-sum Games \\
ICML & Guanghui Qin, Benjamin Van Durme & Nugget: Neural Agglomerative Embeddings of Text \\
ICML & Marc Härkönen, Markus Lange-Hegermann, Bogdan Raiţă & Gaussian Process Priors for Systems of Linear Partial Differential Equations with Constant Coefficients \\
ICML & Xiyao Wang, Wichayaporn Wongkamjan, Furong Huang & Live in the Moment: Learning Dynamics Model Adapted to Evolving Policy \\
ICML & Tanvir Islam, Peter Washington & Personalized Prediction of Recurrent Stress Events Using Self-Supervised Learning on Multimodal Time-Series Data \\
ICML & Krishna Pillutla, Kshitiz Malik, Abdelrahman Mohamed, Michael Rabbat, Maziar Sanjabi, Lin Xiao & Federated Learning with Partial Model Personalization \\
ICML & Jaesik Yoon, Yi-Fu Wu, Heechul Bae, Sungjin Ahn & An Investigation into Pre-Training Object-Centric Representations for Reinforcement Learning \\
ICML & Mehrdad Ghadiri, Matthew Fahrbach, Gang Fu, Vahab Mirrokni & Approximately Optimal Core Shapes for Tensor Decompositions \\
ICML & Arpit Bansal, Ping-yeh Chiang, Michael Curry, Rajiv Jain, Curtis Wigington, Varun Manjunatha, John P Dickerson, Tom Goldstein & Certified Neural Network Watermarks with Randomized Smoothing \\
ICML & Mohamad Amin Mohamadi, Wonho Bae, Danica J. Sutherland & A Fast, Well-Founded Approximation to the Empirical Neural Tangent Kernel \\
ICML & Chuyang Ke, Jean Honorio & Exact Inference in High-order Structured Prediction \\
ICML & Wentao Zhang, Zheyu Lin, Yu Shen, Yang Li, Zhi Yang, Bin Cui & DFG-NAS: Deep and Flexible Graph Neural Architecture Search \\
ICML & Tongzhou Wang, Antonio Torralba, Phillip Isola, Amy Zhang & Optimal Goal-Reaching Reinforcement Learning via Quasimetric Learning \\
ICML & Yi-Fan Zhang, Xue Wang, Kexin Jin, Kun Yuan, Zhang Zhang, Liang Wang, Rong Jin, Tieniu Tan & AdaNPC: Exploring Non-Parametric Classifier for Test-Time Adaptation \\
ICML & Litian Liang, Yaosheng Xu, Stephen McAleer, Dailin Hu, Alexander Ihler, Pieter Abbeel, Roy Fox & Reducing Variance in Temporal-Difference Value Estimation via Ensemble of Deep Networks \\
ICML & Mohammed Nowaz Rabbani Chowdhury, Shuai Zhang, Meng Wang, Sijia Liu, Pin-Yu Chen & Patch-level Routing in Mixture-of-Experts is Provably Sample-efficient for Convolutional Neural Networks \\
ICML & Gal Leibovich, Guy Jacob, Or Avner, Gal Novik, Aviv Tamar & Learning Control by Iterative Inversion \\
ICML & Jiayin Jin, Jiaxiang Ren, Yang Zhou, Lingjuan Lyu, Ji Liu, Dejing Dou & Accelerated Federated Learning with Decoupled Adaptive Optimization \\
ICML & Krzysztof Choromanski, Arijit Sehanobish, Han Lin, Yunfan Zhao, Eli Berger, Tetiana Parshakova, Alvin Pan, David Watkins, Tianyi Zhang, Valerii Likhosherstov, Somnath Basu Roy Chowdhury, Avinava Dubey, Deepali Jain, Tamas Sarlos, Snigdha Chaturvedi, Adrian Weller & Efficient Graph Field Integrators Meet Point Clouds \\
ICML & Simone Parisi, Aravind Rajeswaran, Senthil Purushwalkam, Abhinav Gupta & The Unsurprising Effectiveness of Pre-Trained Vision Models for Control \\
\end{longtable}
\clearpage
\twocolumn

\begin{table*}[t!]
\centering
\begin{tabular}{lp{12cm}}
\toprule
\textbf{Conference} & \multicolumn{1}{c}{\textbf{Paper Title}} \\
\midrule
AAAI & Neuro-symbolic Rule Learning in Real-world Classification Tasks \\
     & Generalization Bounds for Inductive Matrix Completion in Low-noise Settings \\
\addlinespace
\hline
\addlinespace
NeurIPS & A General Framework for Robust G-Invariance in G-Equivariant Networks \\
        & CLIFT: Analysing Natural Distribution Shift on Question Answering Models in Clinical Domain \\
        & Partial Counterfactual Identification of Continuous Outcomes with a Curvature Sensitivity Model \\
        & Attacks on Online Learners: a Teacher-Student Analysis \\
        & Learning Feynman Diagrams using Graph Neural Networks \\
        & Function Classes for Identifiable Nonlinear Independent Component Analysis \\
        & Blackbox Attacks via Surrogate Ensemble Search \\
        & Censored Quantile Regression Neural Networks for Distribution-Free Survival Analysis \\
        & Semi-Discrete Normalizing Flows through Differentiable Tessellation \\
        & Online Decision Mediation \\
        & Exact Generalization Guarantees for (Regularized) Wasserstein Distributionally Robust Models \\
        & Deep Learning with Kernels through RKHM and the Perron-Frobenius Operator \\
        & Bridging RL Theory and Practice with the Effective Horizon \\
        & Reliable learning in challenging environments \\
\addlinespace
\hline
\addlinespace
ICLR & Brain-like representational straightening of natural movies in robust feedforward neural networks \\
     & Broken Neural Scaling Laws \\
     & Parametrizing Product Shape Manifolds by Composite Networks \\
     & Tier Balancing: Towards Dynamic Fairness over Underlying Causal Factors \\
     & Guiding continuous operator learning through Physics-based boundary constraints \\
     & Scaling Laws For Deep Learning Based Image Reconstruction \\
     & Probabilistically Robust Recourse: Navigating the Trade-offs between Costs and Robustness in Algorithmic Recourse \\
     & Domain Adaptation via Minimax Entropy for Real/Bogus Classification of Astronomical Alerts \\
\addlinespace
\hline
\addlinespace
ICML & Why Target Networks Stabilise Temporal Difference Methods \\
     & Nonlinear Advantage: Trained Networks Might Not Be As Complex as You Think \\
     & HyperImpute: Generalized Iterative Imputation with Automatic Model Selection \\
\bottomrule
\end{tabular}
\caption{\textbf{Papers excluded from the analysis.}}
\label{tab:sup_1}
{\footnotesize Note: The papers listed are excluded from the analysis due to tex compilation errors, such as bibtex errors.}
\end{table*}

\end{document}